\documentclass{article}

\usepackage{microtype}
\usepackage{graphicx}
\usepackage{subcaption}
\usepackage{booktabs} 
\usepackage{tabularx}   
\usepackage{ragged2e}   
\usepackage{enumitem}
\usepackage{hyperref}
\usepackage[preprint]{icml2026}

\newcommand{\AACR}{\textit{AACR-Bench}}

\usepackage{multirow} 
\usepackage{amsmath}
\usepackage{amssymb}
\usepackage{mathtools}
\usepackage{amsthm}
\usepackage{booktabs}
\usepackage{graphicx}
\usepackage{listings}
\usepackage[table]{xcolor}

\definecolor{codegreen}{rgb}{0,0.6,0}
\definecolor{codegray}{rgb}{0.5,0.5,0.5}
\definecolor{codepurple}{rgb}{0.58,0,0.82}
\definecolor{backcolour}{rgb}{0.95,0.95,0.92}

\newcommand{\cc}[2]{\cellcolor{orange!#1}#2}

\usepackage[capitalize,noabbrev]{cleveref}

\theoremstyle{plain}

\theoremstyle{definition}

\theoremstyle{remark}

\usepackage[textsize=tiny]{todonotes}

\begin{document}

\twocolumn[
  \icmltitle{AACR-Bench: Evaluating Automatic Code Review with Holistic Repository-Level Context}
  \icmlsetsymbol{equal}{*}

  \begin{icmlauthorlist}
    \icmlauthor{Lei Zhang}{equal,nju}
    \icmlauthor{Yongda Yu}{equal,nju}
    \icmlauthor{Minghui Yu}{nju}
    \icmlauthor{Xinxin Guo}{nju}
    \icmlauthor{Zhengqi Zhuang}{nju}
    \icmlauthor{Guoping Rong}{nju}
    \icmlauthor{Dong Shao}{nju}
    \icmlauthor{Haifeng Shen}{scu}
    \icmlauthor{Hongyu Kuang}{nju}
    \icmlauthor{Zhengfeng Li}{alibaba}
    \icmlauthor{Boge Wang}{alibaba}
    \icmlauthor{Guoan Zhang}{alibaba}
    \icmlauthor{Bangyu Xiang}{alibaba}
    \icmlauthor{Xiaobin Xu}{alibaba}
  \end{icmlauthorlist}

  \icmlaffiliation{nju}{Software Insititute, Nanjing University, Nanjing, China}
  \icmlaffiliation{scu}{Southern Cross University, Gold Coast, Australia}
  \icmlaffiliation{alibaba}{TRE, Alibaba Inc., Hangzhou, China}

  \icmlcorrespondingauthor{Guoping Rong}{ronggp@nju.edu.cn}
  \icmlcorrespondingauthor{Zhengfeng Li}{lizhengfeng.lzf@alibaba-inc.com}
  \icmlkeywords{Machine Learning, ICML}

  \vskip 0.3in
]

\printAffiliationsAndNotice{}

\begin{abstract}

  High-quality evaluation benchmarks are pivotal for deploying Large Language Models (LLMs) in a certain applicaiton field. However, existing benchmarks on Automated Code Review (ACR) suffer from two critical limitations: first, the reliance on noisy, incomplete ground truth derived from raw Pull Request (PR) comments, which constrains the scope of issue detection; second, the lack of 
  multi-language 
  support in repository-level contexts, which restricts the generalizability of evaluation results. To address these challenges, we introduce \AACR, a comprehensive benchmark that provides full cross-file context across multiple programming languages. Unlike traditional datasets, \AACR\ employs an ``AI-assisted, Human expert-verified'' annotation pipeline to uncover latent defects often overlooked in original PRs, resulting in a 285\% increase in issue coverage. Extensive evaluations of mainstream LLMs on \AACR\ reveal that previous assessments may have  either misjudged or only partially captured  model capabilities due to data limitations. Our work establishes a more rigorous standard for ACR evaluation and offers new insights on LLM based ACR, i.e., the granularity/level of context and the choice of retrieval methods significantly impact ACR performance, and this influence varies depending on the LLM, programming language, and the LLM
usage paradigm e.g., whether an Agent architecture is employed. The code, data, and other artifacts of our evaluation set are available at Github\footnote{https://github.com/alibaba/aacr-bench}.
\end{abstract}

\section{Introduction} \label{chap:introduction}

Nowadays, empowered by Large Language Models (LLMs), automated code review(ACR) technologies have been widely studied and adopted~\cite{hou2024large, zhang2023survey}. 
Leveraging the impressive comprehension capabilities of LLMs, researchers have explored various methods for the end-to-end generation of review comments, for example, the  pre-training model schemes proposed by Tufano et al., the fine-tuning strategies that balance cost and performance~\cite{lu2023llama}, and the optimization mechanisms based on reinforcement learning~\cite{le2022coderl}. Recently, Agent-based code review systems have also begun to emerge~\cite{guo2025repoaudit, ren2025hydra}.

The flourishing research and adoption in ACR has intensified the need for credible benchmarks capable of comprehensively evaluating their performance. However, current evaluation frameworks still suffer from critical limitations, which can be summarized into two main issues.

\begin{itemize}
    \item \emph{Incomplete Issue Annotation.}  In general, existing benchmarks 
    directly use raw review comments from real-world historical PRs as the Ground Truth, resulting in inherent limitations regarding issue coverage within the dataset~\cite{liu2025too, khoshnoud2022bugs}. This, in turn, makes it impossible to authentically and faithfully characterize a model's ability to uncover potential issues in a real-world code review scenarios.
    \item \emph{Restricted Context Scope.}
    Many code defects are inherently cross-file, requiring review systems to access repository-level context for accurate detection. Although some existing benchmarks do offer cross-file context awareness \cite{zeng2025benchmarking, guo2025codefuse}, they are predominantly limited to a single programming language, such as Python. This language-specific focus not only constrains the generalizability of evaluation findings but may also introduce structural biases tied to the linguistic characteristics of that language. Consequently, such benchmarks fail to adequately represent the multilingual reality of contemporary software development.

\end{itemize}
 These deficiencies collectively undermine the credibility and scope of current ACR evaluations, highlighting the necessity for a more comprehensive and crediable benchmark.

To address the aforementioned challenges, we propose \AACR—a multilingual ACR benchmark that supports Repository-level Context-awareness. First, to address the limited issue annotation in existing datasets, we constructed a high-quality hybrid corpus. Beyond collecting 391 real-world review comments from GitHub PRs, we incorporated extensive manual issue annotation: 80 senior software engineers (each with 2+ years of industry experience) meticulously reviewed 2,145 comments generated by two ACR systems across six LLMs. This annotation strategy substantially improves issue coverage and enables \AACR\ to more accurately and comprehensively assess a model's capability to uncover latent issues compared to conventional datasets.
Secondly, 
to address the issues of context and language diversity, \AACR\ provides complete repository-level dependency information while broadly covering 10 mainstream programming languages. The full \AACR\ dataset can be retrieved in the supplementary materials of this paper and will be open-sourced later on.

Our main contributions are summarized as follows:
\begin{itemize}[topsep=0pt,itemsep=2pt,parsep=0pt,leftmargin=*]
    \item We combine multi-model generation and large-scale human annotation to improve issue coverage so as to establish a better groundtruth dataset.
    \item We propose \AACR, the first multilingual, repository-level context-aware benchmark for LLM enabled ACR tasks.
    \item We conduct a comprehensive empirical evaluation of the  mainstream LLMs with \AACR\ and reveal new insights towards mainstream LLM's ACR capabilities.
\end{itemize}

\section{Related Work}
We position this study by examining two key dimensions: the current state of large language model (LLM)-enabled ACR, and the evolving landscape of corresponding specialized benchmark methodologies.

\subsection{Automatic Code Review Based on LLM}

Recent advances in LLMs have driven growing research in automated code review (ACR), with methods spanning several categories. Early work includes T5-based ACR~\cite{tufano2022using} and CodeReviewer~\cite{li2022automating}, pretrained on large diff–comment pairs. To improve efficiency, Llama-Reviewer~\cite{lu2023llama} employed Parameter-Efficient Fine-Tuning, while Sun et al. introduced a ``Data Flywheel” for continuous model evolution via iterative feedback~\cite{sun2025bitsai, yu2024fine}.
Recent efforts align models with human preferences: Yu et al.~\cite{yu2024distilling} used Kahneman–Tversky Optimization to enhance review utility, and Kapadnis et al.~\cite{kapadnis2025crscore++} combined static analysis with Direct Preference Optimization to better detect latent defects.
To address limited context, Zhang et al.~\cite{zhang2025laura} applied Retrieval-Augmented Generation to incorporate project-level semantics. Further extending reasoning depth, agent-based frameworks employ multi-agent collaboration~\cite{ren2025hydra, li2025issue, sharanarthi2025multi} and dialectical interactions such as debate~\cite{tang2024codeagent} to produce more comprehensive and objective reviews.

\subsection{Code Review Benchmarks}
Benchmarks are essential for defining the capability boundaries of LLMs and guiding algorithmic evolution~\cite{jimenez2023swe}. In ACR, the rapid emergence of diverse methods underscores the need for a standardized benchmark to systematically assess performance and guide future optimization. Currently, ACR benchmarking remains nascent, with efforts falling into two categories: accompanying datasets from individual studies and dedicated evaluation datasets.
The most widely used accompanying dataset is CodeReviewer~\cite{li2022automating}, employed in many recent works~\cite{lu2023llama,yu2024fine,yu2024distilling,kapadnis2025crscore++,ren2025hydra,li2025issue}. However, it provides only diff-level code fragments and lacks repository-level context, which limits its realism relative to production environments. A few studies~\cite{sun2025bitsai,sharanarthi2025multi} forgo static datasets altogether, relying solely on live user feedback.
Dedicated benchmarks include SWR-Bench~\cite{zeng2025benchmarking} (derived from SWE-Bench’s 12 Python projects) and CodeFuse-CR-Bench~\cite{guo2025codefuse} (70 Python projects). While both offer full repository context, their exclusive focus on Python limits generalizability. ContextCRBench~\cite{hu2025benchmarking} improves language diversity (9 languages across 90 repositories) but restricts context to the file level, preventing evaluation of cross-file(i.e., repo-level) reasoning needed for complex defects.

Motivated by the aforementioned landscape, this paper introduces the first multi-language with  repository-level context ACR evaluation dataset, i.e, \AACR, aiming to establish a benchmark that more closely reflects real-world production environments.

\section{The AACR-Bench}

\subsection{Overview of AACR-Bench}

\begin{figure*}
    \centering
    \includegraphics[width=0.95\linewidth]{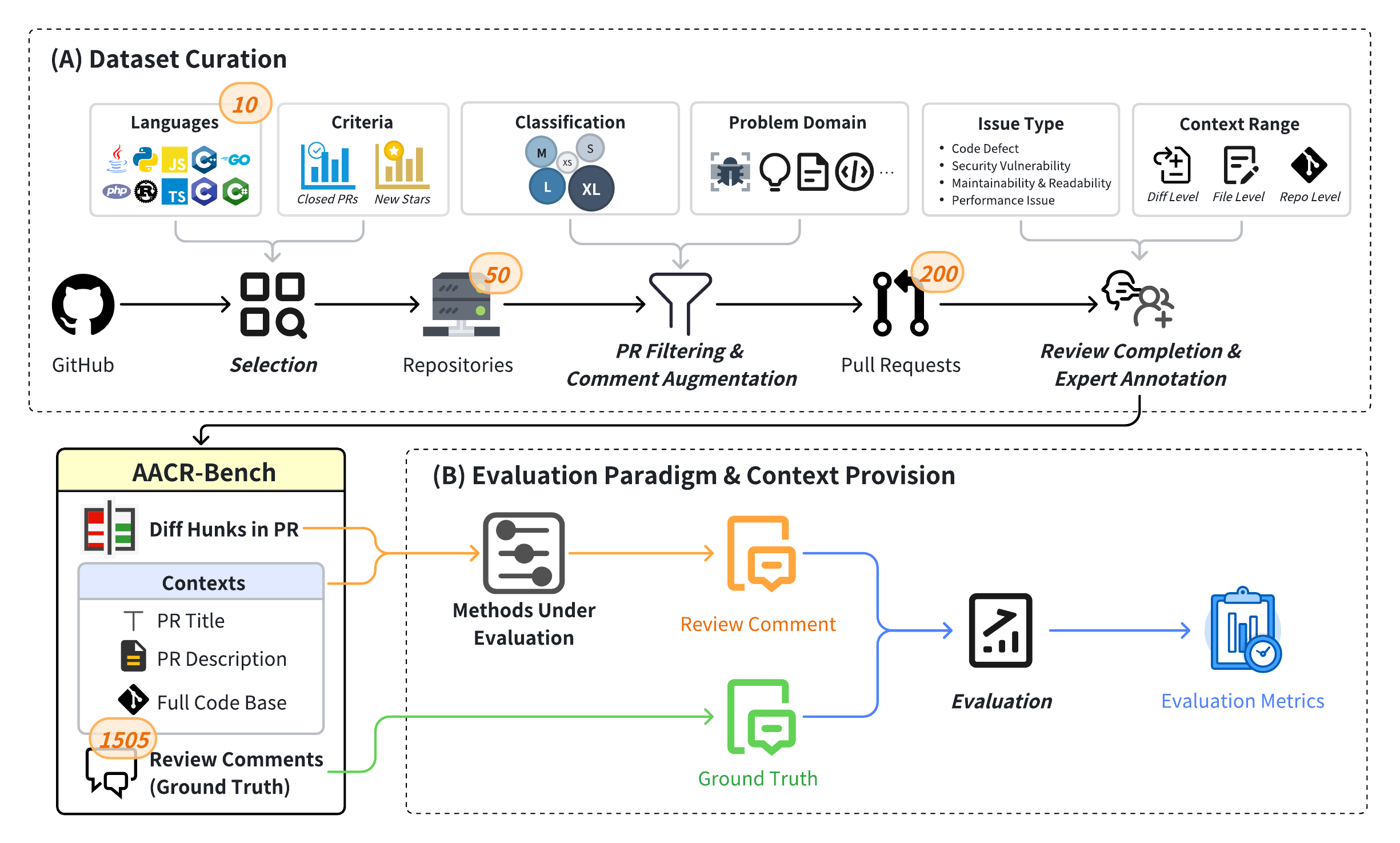}
    \caption{\textbf{Overview of AACR-Bench.} \AACR\ contains $200$ PRs and $1,505$ fine-grained review comments extracted and curated from $50$ popular repositories, covering $10$ mainstream programming languages.}
    \label{fig:aacr-overview}
\end{figure*}

\AACR\ is a repository-level benchmark tailored for evaluating the end-to-end performance of ACR methods/systems. Figure~\ref{fig:aacr-overview} show the overview of \AACR. In general, \AACR\ contains $200$ PRs and $1,505$ fine-grained review comments extracted and curated from $50$ popular repositories, covering $10$ mainstream programming languages. Besides, \AACR\ incorporates a hybrid of model-augmented human reviews and fully model-generated reviews, all of which have been rigorously verified through human expert annotation to ensure benchmark credibility.

\begin{table}[htb]
    \centering
    \scriptsize
    \caption{Context Range Categories of Review Comments}
    \label{tab:comment_context}
    \begin{tabular}{c p{18em} c}
        \toprule
        \textbf{Context level} & \textbf{Description} & \textbf{\# Comment} \\
        \midrule
        Diff  & Review comments that can be provided by looking solely at the current diff hunk & 754 \\
        File  & Review comments that require the context of the entire file containing the diff hunk & 518 \\
        Repo  & Review comments that require repository-wide context, including PR metadata and the contents of other files in the code base & 233 \\
        \bottomrule
    \end{tabular}
\end{table}

To align with the diff-oriented paradigm of modern code review, \AACR\ designates each PR as an evaluation unit. During execution, the ACR method under assessment traverses all \emph{diff} Hunks within a PR to generate review comments. Performance is evaluated by matching these generated comments against the Ground Truth (comprising 1,505 items) and calculating specific accuracy metrics(e.g., \emph{Precision}, \emph{Recall} and \emph{F1-score}.)

\subsection{Dataset Curation}

\begin{figure}[ht]
    \centering
    \includegraphics[width=0.75\linewidth]{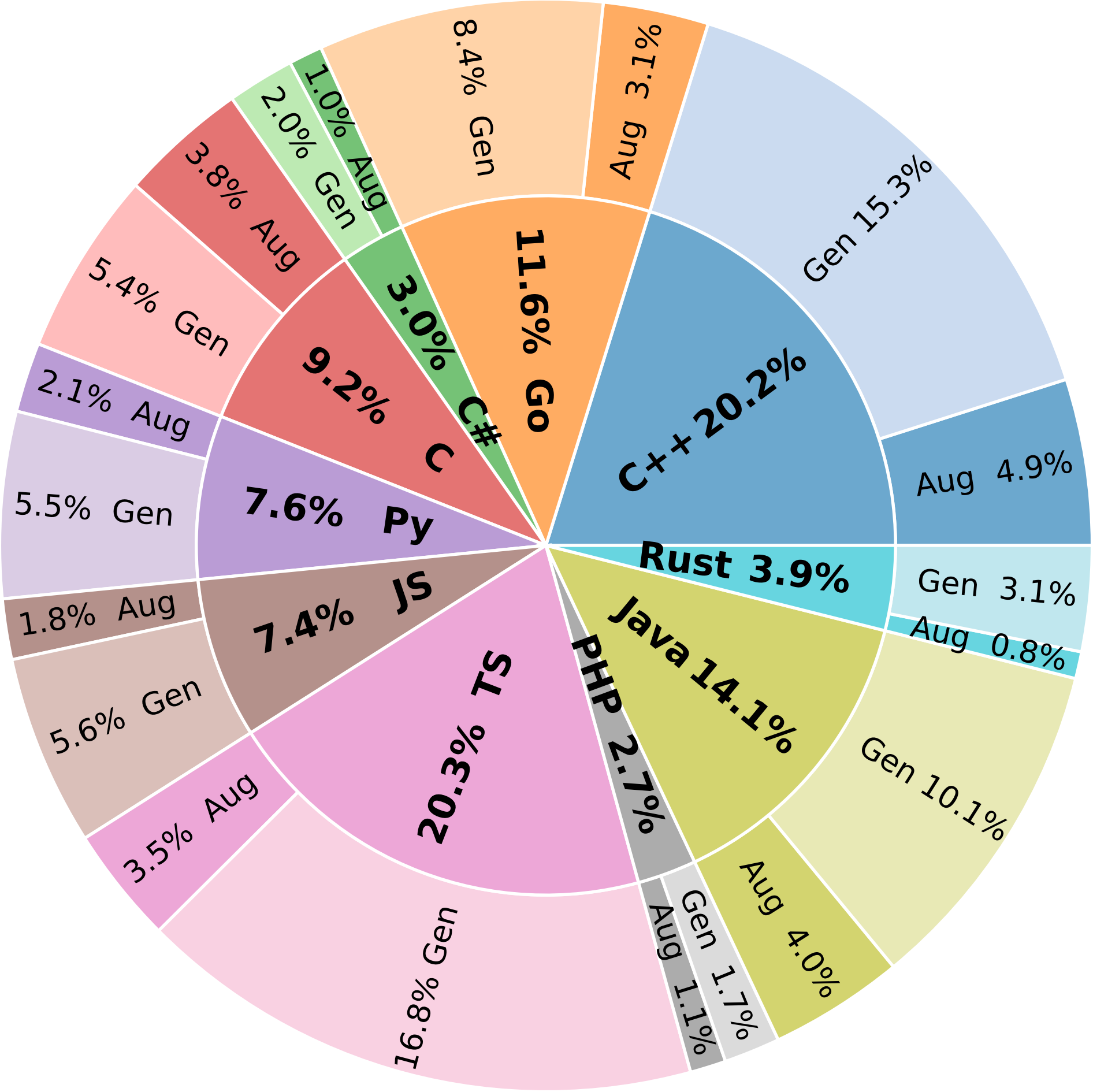}
    \caption{\textbf{Distribution of Review Comments in \AACR\ .} TS, JS, and Py stand for TypeScript, JavaScript, and Python, respectively. ``Aug'' denotes comments augmented from original PR reviews, while ``Gen'' denotes comments generated by the 6 LLMs.}
    \label{fig:comment-distribution}
\end{figure}

As discussed in Section~\ref{chap:introduction}, existing benchmarks suffer from single-language confinement and incomplete ground truth derived solely from historical PR data. To bridge these gaps, we constructed \AACR\, a multilingual, repository-level benchmark. Core considerations and treatments are listed as below:

\paragraph{Selection of programming languages and repositories} To ensure both the currency and diversity of our evaluation benchmark, we selected the top 10 programming languages based on the StackOverflow Developer Survey 2025\footnote{https://survey.stackoverflow.co/2025/technologyÅ}, namely: \textit{JavaScript}, \textit{Python}, \textit{TypeScript}, \textit{Java}, \textit{C\#}, \textit{C++}, \textit{C}, \textit{PHP}, \textit{Go}, and \textit{Rust}. For each language, we identified candidate repositories on GitHub that ranked within the top 2,000 in terms of both newly acquired stars and closed PRs from December 1, 2024, to December 1, 2025. From this candidate pool, we selected the top 5 repositories with the highest number of new stars. Consequently, our final dataset comprises 50 repositories across these 10 distinct languages (5 repositories per language), serving as the original source for extracting PRs and review comments.

\paragraph{PR filtering and comment augmentation}
With the extracted PRs from 50 selected repositories, we discard comments that did not identify explicit issues. To ensure data quality, we applied five filtering criteria: (1) PR titles and descriptions must be in English; (2) changed lines of code must be $\le 1,000$ (aligning with Google's code review practice for effective review~\cite{solmaz2025google}); (3) the primary programming language of modified files must match the repository's main language; (4) a PR must contain $> 2$ inline comments, including at least one adopted comment that led to code modification; (5) changes detached from the project's business context or lacking semantic meaning were excluded.
Finally, to ensure the representativeness and diversity of the benchmark, we applied stratified sampling to the filtered candidates based on the repository, PR problem domain~\cite{jimenez2023swe,guo2025codefuse}, and change size, constructing a core dataset of 200 PRs. As code review often involves multi-turn dialogue, directly using raw comments may lose context or introduce noise. Focusing on the revision with the most inline comments per PR, we employed an LLM to perform deep semantic analysis on review threads. This allowed us to extract confirmed code defects from multi-turn interactions and reorganize them into ``Augmented Review Comments.''

\paragraph{Review completion and expert annotation} To address the Label Incompleteness stemming from the often under-reviewed nature of GitHub PRs~\cite{bacchelli2013expectations}, we leveraged LLMs to comprehensively supplement the review comments for each PR. We constructed a generation matrix comprising 6 mainstream open- and closed-source models(Claude-4.5-Sonnet, Qwen3-Coder-480B-A35B-Instruct~\cite{qwen3technicalreport}, GPT-5.2, Deepseek-V3.2~\cite{deepseekai2025deepseekv32}, GLM-4.7~\cite{5team2025glm45agenticreasoningcoding}, Gemini-3-Pro) to mitigate single-model bias and ensure output diversity. Review comments were generated in parallel through two heterogeneous frameworks: an internal review system and the open-source agent, Claude Code. Following semantic de-duplication, the generated comments were merged with the augmented human reviews to form a candidate set for human verification later on. The collection was subjected to rigorous human annotation by 80 senior software engineers, each with over two years of professional experience. Every comment was annotated by at least two annotators independently, with any discrepancies resolved through discussion by a six-member core team. Annotators validated the correctness of the comments and categorized the issue types\cite{sun2025bitsai}. Distinguishing our work from existing benchmarks, we innovatively annotated the context level required to formulate each comment, enabling the assessment of detection difficulty across varying context dependencies. This step yielded a total of 1,505 review comments, including 391 augmented from original reviews and 1,114 augmented by LLMs and human experts, representing a 285\% increase in issue coverage. 
Figure~\ref{fig:comment-distribution} illustrates the language distribution and the proportion of augmented human versus model-generated comments.

\subsection{Comparisions with Existing Datasets}

\begin{table}[]
    \centering
    \caption{Comparisons between Existing Benchmarks and Ours}
    \label{tab:benchmark_comparisons}
    \scriptsize
    \begin{tabular}{p{7em}p{4em}p{5em}p{5em}p{5em}}
        \toprule
        \textbf{Dataset} & \textbf{Multi-Languages support}& \textbf{Repo-level context support} & \textbf{Source of review comments} & \textbf{Context scope annotation} \\
        \midrule
        CodeFuse-CR-Bench
        & no & yes & Raw data & no \\
        SWR-Bench
        & no & yes & Raw data & no \\
        ContextCRBench
        & yes & no & Raw data & no \\
        \AACR\ (ours) & yes & yes  & LLM\&Expert Augmented & yes \\
        \bottomrule
    \end{tabular}
\end{table}

A brief comparison between existing ACR benchmarks and ours are shown in Table~\ref{tab:benchmark_comparisons}. Our benchmark distinguishes itself as the first to provide repository-level context within a multi-language environment, synthesizing the strengths of existing works. While Codefuse-CR-Bench~\cite{guo2025codefuse} and SWR-Bench~\cite{zeng2025benchmarking} offer repository contexts, they are restricted to Python; conversely, multi-language datasets like ContextCRBench~\cite{hu2025benchmarking} and CodeReview~\cite{li2022automating} are limited to file-, function- or even diff-level contexts. We bridge this gap by evaluating tools across 10 popular programming languages with comprehensive context, including PR metadata and complete code repositories.

Furthermore, we address the noise and incompleteness inherent in raw GitHub data used by prior benchmarks~\cite{guo2025codefuse,zeng2025benchmarking,hu2025benchmarking,li2022automating}. Instead of relying on raw extraction, we employ a rigorous construction methodology: we use LLMs to augment technical issues from human inputs and expand defect coverage using six advanced models (including Qwen3-Coder-480B, GPT-5.2, and Gemini-3-Pro). To ensure reliability, all comments—whether augmented or generated—are meticulously verified by a large-scale team consisting of 80 professional software engineers.

Finally, beyond standard PR and comment labels~\cite{guo2025codefuse,zeng2025benchmarking,hu2025benchmarking}, we innovatively annotate the required context scope for each review comment. This novel dimension enables a granular analysis of an ACR method's capacity to detect issues dependent on varying context depths and scope.

\section{Experiments}
Our experiments are designed to validate impacts derived from the core novel features of the proposed dataset(\AACR), i.e, multi-language support, repository-level context, and a more comprehensive level of defect exposure.

\subsection{Evaluation Settings}

\textbf{Models} We selected the latest models from mainstream open-source providers and the most recent versions of major commercial large models. As the result, our experiment comprises three open-source models (Qwen3-Coder-480B-A35B-Instruct~\cite{qwen3technicalreport}, DeepSeek-V3.2~\cite{deepseekai2025deepseekv32}, GLM-4.7~\cite{5team2025glm45agenticreasoningcoding}) and two commercial models (GPT-5.2, Claude-4.5-Sonet). We evaluated all selected models using our complete dataset. For simplicity, Qwen3-Coder-480B-A35B-Instruct will be called Qwen-480B-Coder in the following contents.

\textbf{Context Retrival Method}
One distinctive feature of our dataset is the provision of repository-level contextual dependencies. However, different retrieval methods also significantly influence the model’s reasoning performance~\cite{zhang2023repocoder, liu2023repobench}. In this experiment, the following methods were adopted:
\begin{itemize}[topsep=0pt,itemsep=1pt,parsep=0pt,leftmargin=*]
    \item \textbf{No context}: Serves as a comparative basis.
    \item \textbf{BM25}: A classic text similarity method~\cite{robertson2009probabilistic}, also applied in a previous similar study~\cite{guo2025codefuse}.
    \item \textbf{Embedding}: As one of the current state-of-the-art (SOTA) models, Qwen3-Embedding-8B is adopted for vector similarity retrieval.
    \item \textbf{Agent-based approach}: We selected Claude Code, a widely-used agent framework that supports code review.
\end{itemize}

Additionally, it should be noted that for similarity-based retrieval methods, the number of retrieved code contexts was uniformly set to 3.  In contrast, the Agent method allows the Claude Code framework to autonomously decide the number of contexts to retrieve. And in all cases (even with the \textbf{No context} approach, the exactly same PR title and description were provided.

\paragraph{Metrics} 
We evaluate the performance of various ACR approaches against \AACR\ using the common three matrics, i.e., $\mbox{\it Precision}$, $\mbox{\it Recall}$ and $\mbox{\it F1-score}$.

\begin{table*}[htb]
  \centering
  \caption{Performance Comparison (\%) of Models Across Different ACR Methods}
  \label{tab:model_performance}
  \renewcommand{\arraystretch}{1.2}
  \resizebox{0.7\textwidth}{!}{
  \begin{tabular}{llcccc}
    \toprule
    \textbf{Method} & \textbf{Model} & \textbf{Avg Comments} & \textbf{Recall (\%)} & \textbf{Precision (\%)} & \textbf{F1 (\%)} \\
    \midrule
    \multirow{5}{*}{\textbf{Agent}} 
      & Claude-4.5-Sonnet & \cc{4}{0.0890} & \cc{21}{10.10} & \cc{100}{39.90} & \cc{100}{16.12} \\
      & Deepseek-V3.2     & \cc{6}{0.1525} & \cc{10}{4.78}  & \cc{28}{11.00} & \cc{41}{6.67} \\
      & GLM-4.7           & \cc{6}{0.1448} & \cc{10}{4.72}  & \cc{29}{11.50} & \cc{42}{6.69} \\
      & GPT-5.2           & \cc{4}{0.1063} & \cc{6}{2.99}  & \cc{25}{9.90}  & \cc{28}{4.59} \\
      & Qwen-480B-Coder   & \cc{4}{0.1004} & \cc{9}{4.39}  & \cc{38}{15.30} & \cc{42}{6.82} \\
    \midrule
    \multirow{5}{*}{\textbf{Embedding}} 
      & Claude-4.5-Sonnet & \cc{75}{1.8902} & \cc{91}{42.86} & \cc{20}{8.00}  & \cc{84}{13.48} \\
      & Deepseek-V3.2     & \cc{100}{2.5230} & \cc{77}{36.35} & \cc{13}{5.10}  & \cc{55}{8.94} \\
      & GLM-4.7           & \cc{034}{0.8657} & \cc{57}{26.98} & \cc{28}{11.00} & \cc{97}{15.63} \\
      & GPT-5.2           & \cc{98}{2.4709} & \cc{100}{47.24} & \cc{17}{6.70}  & \cc{73}{11.74} \\
      & Qwen-480B-Coder   & \cc{33}{0.8335} & \cc{51}{24.25} & \cc{26}{10.20} & \cc{89}{14.36} \\
    \midrule
    \multirow{5}{*}{\textbf{No context}} 
      & Claude-4.5-Sonnet & \cc{68}{1.7220} & \cc{91}{42.86} & \cc{22}{8.70}  & \cc{90}{14.46} \\
      & Deepseek-V3.2     & \cc{91}{2.2875} & \cc{77}{36.54} & \cc{14}{5.60}  & \cc{60}{9.71} \\
      & GLM-4.7           & \cc{34}{0.8573} & \cc{58}{27.57} & \cc{28}{11.30} & \cc{99}{16.03} \\
      & GPT-5.2           & \cc{93}{2.3537} & \cc{100}{47.11} & \cc{18}{7.00}  & \cc{76}{12.19} \\
      & Qwen-480B-Coder   & \cc{40}{1.0217} & \cc{58}{27.44} & \cc{24}{9.40}  & \cc{87}{14.00} \\
    \midrule
    \multirow{5}{*}{\textbf{BM25}} 
      & Claude-4.5-Sonnet & \cc{86}{2.1698} & \cc{76}{35.75} & \cc{15}{5.80}  & \cc{62}{9.98} \\
      & Deepseek-V3.2     & \cc{35}{0.8851} & \cc{58}{27.38} & \cc{27}{10.90} & \cc{97}{15.59} \\
      & GLM-4.7           & \cc{36}{0.9070} & \cc{56}{26.25} & \cc{26}{10.20} & \cc{91}{14.69} \\
      & GPT-5.2           & \cc{69}{1.7461} & \cc{92}{43.59} & \cc{22}{8.80}  & \cc{91}{14.64} \\
      & Qwen-480B-Coder   & \cc{95}{2.3906} & \cc{97}{45.85} & \cc{17}{6.70}  & \cc{73}{11.69} \\
    \bottomrule
  \end{tabular}
  }
\end{table*}

\subsection{Main Results}

 This section presents a comprehensive evaluation of diverse ACR methodologies and models on our benchmark. We specifically analyze the impact of context retrieval, model architectures, and agent frameworks on the overall quality of generated reviews. It is important to note that both the current practical applications of ACR~\cite{sadowski2015tricorder, distefano2019scaling} and prior research~\cite{tao2012software} consistently highlight the critical role of contextual dependency information in ACR. Therefore, all experiments in this paper are based on the assumption that this represents the geneunin paradigm established for ACR. In short, unless serving as a comparative basis, we do not evaluate the performance of LLMs on ACR tasks without contextual information.

\subsubsection{ACR performance benchmarked with a more sufficient exposure to defects}

As evidenced by the experimental data in Table~\ref{tab:model_performance}, Agent-based methods exhibit review performance that are distinctly divergent from non-Agent baselines. Meanwhile, introducing contextual information does not always yield positive gains in ACR task performance.

\paragraph{Comparison between Agent-based and traditional approaches}

Agent-based methods (e.g., Claude Code) consistently generate substantially fewer review comments per patch ($0.08 \sim 0.15$) compared to traditional approaches. For instance, Claude-4.5-Sonnet in Agent mode attains a Precision of $39.90\%$, far exceeding its $8.70\%$ in ``No context'' mode, yet accompanied by a notably lower Recall of $10.10\%$. This indicates that while Agents excel in precision-critical scenarios, they may overlook many potential defects—a tendency attributable to the ``contextual tunnel vision'' effect induced by their focused retrieval mechanism~\cite{liu2023lost}. In contrast, traditional methods tend to produce an excess of comments (e.g., GPT-5.2 averages $2.47$ per patch), which can obscure meaningful insights amidst noise.
Furthermore, the effectiveness of Agent-based methods exhibits strong model-specific dependency, a sensitivity not observed with other retrieval strategies. Although Claude-4.5-Sonnet achieves exceptional precision in Agent mode, other models do not show similar gains and may even experience performance degradation. For example, GPT-5.2, despite strong baseline performance in ``no context'' mode, suffers a sharp decline in Agent mode, with Precision falling to $9.90\%$ and Recall dropping to $2.99\%$. This contrast suggests that general-purpose model capabilities do not directly translate to proficiency in Agent-based ACR tasks~\cite{liu2023agentbench}.

\paragraph{Impact of context retrieval methods}
Table~\ref{tab:model_performance} further indicates that context retrieval is not universally beneficial. For models with strong inherent reasoning, naive RAG can introduce harmful noise, while various open-source models show distinct preferences for specific retrieval strategies.
For instance, Claude-4.5-Sonnet performs robustly in ``no context'' mode (F1=$14.46$). However, providing top-3 contexts via BM25 causes performance to drop sharply (F1=$9.98$, a $31\%$ decrease; Precision falls from $8.70\%$  to $5.80\%$). Embedding-based retrieval similarly degrades its results.
Different models favor different strategies. DeepSeek-V3.2 achieves its best performance under BM25: Avg. Comments per patch decreases from 2.29 (``No context”) to $0.89$, while Precision jumps from $5.10\%$ to $10.90\%$, yielding an F1 of $15.59$. Conversely, Qwen-480B-Coder peaks using Embedding-based retrieval (F1=$14.36$), significantly outperforming its BM25 score ($11.69$).
Thus, no single retrieval method is optimal for all models; performance is highly dependent on the specific combination of model and retrieval mode.

\paragraph{Key observation}
\emph{The Agent-based approach consistently produces far fewer review comments than other methods, and its effectiveness is strongly model-dependent. Performance varies markedly across different LLMs under this paradigm. Furthermore, different models respond differently to context retrieval methods, e.g., 
pairing Claude-4.5-Sonnet with Agent frameworks, DeepSeek with BM25, and Qwen with Embedding consistently present optimal performance.}

\begin{table}[htbp]
  \centering
  \footnotesize
  \caption{Performance on Finding Issues Using Different Contextual Scopes Across Various ACR Methods (Recall only)}
  \label{tab:context_recall}
  \renewcommand{\arraystretch}{1.1} 
  \begin{tabular}{llccc}
    \toprule
    \multirow{2}{*}{\textbf{Method}} & \multirow{2}{*}{\textbf{Model}} & \multicolumn{3}{c}{\textbf{Context \&  Recall (\%)}} \\
    \cmidrule(lr){3-5}
          &       & \textbf{Diff} & \textbf{File} & \textbf{Repo} \\
    \midrule

    \multirow{5}{*}{\textbf{Agent}} 
          & Claude-4.5-Sonnet & \cc{23}{11.95} & \cc{38}{16.84} & \cc{32}{13.77} \\
          & Deepseek-V3.2     & \cc{8}{4.28}   & \cc{11}{4.64}  & \cc{19}{8.00} \\
          & GLM-4.7           & \cc{12}{6.50}  & \cc{14}{5.95}  & \cc{10}{4.40} \\
          & GPT-5.2           & \cc{6}{3.21}   & \cc{6}{2.52}   & \cc{8}{3.45} \\
          & Qwen-480B-Coder   & \cc{9}{4.49}   & \cc{12}{5.24}  & \cc{14}{5.94} \\
    \midrule
    \multirow{5}{*}{\textbf{No context}} 
          & Claude-4.5-Sonnet & \cc{87}{45.76} & \cc{93}{40.54} & \cc{90}{38.63} \\
          & Deepseek-V3.2     & \cc{75}{39.26} & \cc{74}{32.43} & \cc{86}{36.91} \\
          & GLM-4.7           & \cc{60}{31.43} & \cc{57}{24.90} & \cc{49}{21.03} \\
          & GPT-5.2           & \cc{97}{50.66} & \cc{100}{43.82}& \cc{100}{42.92} \\
          & Qwen-480B-Coder   & \cc{65}{33.82} & \cc{52}{22.59} & \cc{41}{17.60} \\
    \midrule
    \multirow{5}{*}{\textbf{BM25}} 
          & Claude-4.5-Sonnet & \cc{89}{46.68} & \cc{94}{41.31} & \cc{90}{38.63} \\
          & Deepseek-V3.2     & \cc{73}{38.20} & \cc{75}{32.82} & \cc{80}{34.33} \\
          & GLM-4.7           & \cc{57}{29.71} & \cc{63}{27.61} & \cc{45}{19.31} \\
          & GPT-5.2           & \cc{93}{48.94} & \cc{100}{43.82}& \cc{94}{40.34} \\
          & Qwen-480B-Coder   & \cc{57}{29.97} & \cc{55}{23.94} & \cc{45}{19.31} \\
    \midrule
    \multirow{5}{*}{\textbf{Embedding}} 
          & Claude-4.5-Sonnet & \cc{89}{46.82} & \cc{87}{38.03} & \cc{95}{40.77} \\
          & Deepseek-V3.2     & \cc{75}{39.39} & \cc{76}{33.40} & \cc{77}{33.05} \\
          & GLM-4.7           & \cc{57}{30.11} & \cc{61}{26.64} & \cc{41}{17.60} \\
          & GPT-5.2           & \cc{100}{52.39}& \cc{96}{42.28} & \cc{97}{41.63} \\
          & Qwen-480B-Coder   & \cc{52}{27.06} & \cc{52}{22.78} & \cc{43}{18.45} \\
    \bottomrule
  \end{tabular}
\end{table}

\subsubsection{Context level-wise impact on ACR performance}

We evaluate the efficacy of various approaches in identifying issues that necessitate varying  context levels(defined in Table~\ref{tab:comment_context}). Since it is impossible to ascertain the precise context level implicitly utilized in each inference, we exclusively report Recall metrics.

\paragraph{The impacts derived from contextual levels}
As presented in Table~\ref{tab:context_recall}, 
with the notable exception of Agent frameworks, all context retrieval methodologies (No context, BM25, Embedding) exhibit a distinct trend of performance decay across varying context levels, consistently adhering to the hierarchy of $\text{Diff} > \text{File} > \text{Repo}$. Taking the ``No context'' method as an exemplar: Qwen-480B-Coder's performance degrades from $33.82\%$ at the Diff level to $22.59\%$ at the Repo level, further declining to $17.60\%$ at the Repo level. Similarly, GPT-5.2 demonstrates a parallel downward trajectory.
Even the incorporation of retrieval augmentation (BM25 or Embedding) fails to reverse this phenomenon of performance attenuation correlated with expanding contextual scope.
In stark contrast, Agent frameworks (implemented via Claude Code) generally exhibit an inverse trend, frequently outperforming in complex Repo-level scenarios compared to isolated \emph{diff} scenarios.
For instance, under the Agent framework, DeepSeek-V3.2's performance improved from $4.28\%$ at the \emph{diff} level to $8.00\%$ at the Repo level; similarly, Qwen-480B-Coder ascended from $4.49\%$ to $5.94\%$. While this suggests that Agents can leverage multi-turn interactions to effectively retrieve complex contextual information, the extremely low scores at the \emph{diff} level (e.g., DeepSeek's $4.28\%$ VS. ``No context'' performance of $39.26\%$) may imply Agents may become preoccupied with external dependencies, thereby overlooking conspicuous local issues inherent within the \emph{diff} itself.

\paragraph{Key observation} \emph{ACR methods other than the Agent-based approach show a declining ability to detect issues as the required context level/scope increases, whereas the Agent approach exhibits the opposite trend.}

\subsubsection{Languange-wise impact on ACR performance} 
\begin{figure*}[h]
    \centering
    \includegraphics[width=1.05\linewidth]{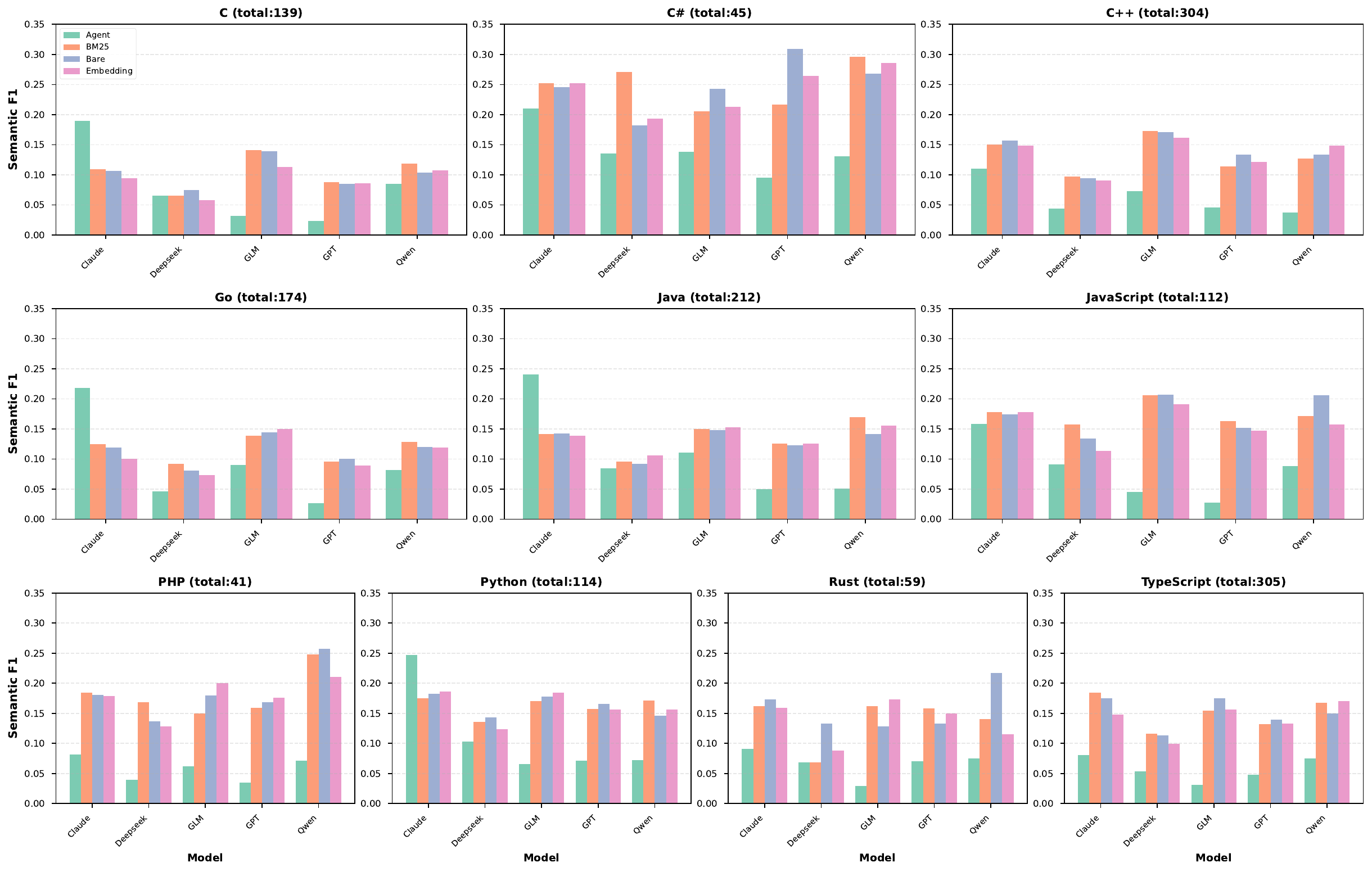}
    \caption{Language-wise Code Review Performance}
    \label{fig:performence_by_lan}
\end{figure*}

Figure~\ref{fig:performence_by_lan} reveals that programming languages may also impact the performance of various ACR approaches, which to a fair degree confirm the importance of expanding the benchmark to multiple programming languages. 

\paragraph{Language-specific bias in model performance} 
The effectiveness of different ACR methods varies considerably across programming languages, revealing strong language-specific bias. For instance, Claude-4.5-Sonnet—the best-performing model in Agent tasks—exhibits a clear performance hierarchy: it achieves notably higher F1 scores on Python (0.247), Java (0.241), Go (0.218), and C (0.189), forming a distinct top tier. In contrast, its performance drops sharply on TypeScript (0.081), PHP (0.082), and Rust (0.091), with gaps as large as 3× between Python and TypeScript.
We attribute this discrepancy primarily to the uneven distribution of the models' training data. Languages such as Python and Java, with their extensive ecosystems and abundance of high-quality open-source repositories, likely provide sufficient and well-curated training corpora. In contrast, for languages like Rust, the relatively scarce corpus often hinders LLMs/agents from acquiring the necessary knowledge to execute accurate planning steps.

Meanwhile, 
a cross-comparison of all mainstream frameworks (No context, BM25, Embedding, Agent) also reveals that C\# exhibits exceptional interpretability and model-friendliness, whereas the C language consistently remains at the bottom of the performance spectrum. In the context-free ``No context'' mode, GPT-5.2 achieved the highest F1 score across the board (0.309) on C\#, while the same model scored only 0.085 on C. This trend is further corroborated by Qwen-480B-Coder (C\# $0.268$ vs. C $0.104$).  
Given the application status of both languages, we can reasonably assume that C\# and C  provide comparable abundance for training corpora.
In this sense, this phenomenon deeply reflects the impact of programming language intrinsic characteristics on the performance of LLMs. For example, 
as a strongly typed, and object-oriented language, C\# possesses rigorous namespace management and explicit type definitions. In contrast, C  relies heavily on pointer operations, macro definitions, and implicit memory management. Its dependency information is often implicit within unstructured header files or linking logic. We hypothesize that such structural features exert a substantial influence on the ACR performance of LLMs.

\paragraph{Performance impact of context across languages} 
While context is generally expected to improve model performance, our experiments reveal a counter-intuitive trend: introducing context leads to significant ``Contextual Backwardness'' in most languages, with only a few demonstrating robustness. Comparing the ``No context'' mode with the ``Agent/Retrieval'' modes reveals two distinct behavioral patterns.
For C\#, C++, JavaScript, PHP, Python, Rust, and TypeScript, the introduction of complex context frameworks paradoxically introduced noise. Taking GPT-5.2 as an example, its performance on C\# plummeted from 0.309 in ``no context'' mode to 0.095 in Agent mode; similarly, Python declined from 0.165 to 0.071. This indicates that for these languages, externally retrieved context or redundant planning steps generated by the Agent severely interfered with the model's intrinsic judgment.
In contrast, only C, Go, and Java maintained stability or even achieved improvements within complex contexts. The most prominent instance is Claude-4.5-Sonnet, which realized substantial performance leaps via the Agent framework on Go ($0.120 \rightarrow 0.218$), Java ($0.142 \rightarrow 0.241$), and C ($0.106 \rightarrow 0.189$). 
This suggests that when considering whether to provide contextual information to models to improve their performance in ACR, the answer varies entirely depending on the programming language.

\vspace{-0.35cm}

\paragraph{Key observation} \emph{Different ACR methods exhibit significant language-specific bias. For certain languages, introducing context retrieval methods may even degrade performance.}

\section{Error Analysis}

Detailed case studies of incorrect review comments generated by the model are provided in the Appendix~\ref{app:case-study}. We observe that current models continue to suffer from knowledge errors during code review. Furthermore, noisy data introduced by context retrieval is identified as a significant factor contributing to the generation of incorrect review comments.

\section{Conclusion}

We proposed \AACR, a  multi-lingual benchmark designed to evaluate ACR systems with repository-level context. \AACR\ fills a critical gap in the field by better measuring the ability of various  methods to leverage complex, repo-level context to perform ACR.
Our extensive empirical evaluation on \AACR\ reveals that:

\emph{The granularity/level of context and the choice of retrieval methods significantly impact ACR performance, and this influence varies depending on the LLM, programming language, and the LLM usage paradigm (e.g., whether an Agent architecture is employed). } 

As long as the above finding strongly underscores the necessity of constructing a dataset such as\AACR,  our work also has limitations. Specifically, although we leveraged LLM-generated reviews to augment our dataset, constructing a fully comprehensive Ground Truth remains a formidable challenge due to the inherent complexity and subjectivity of real-world software systems. Future work will focus on further expanding the dataset scale and exploring more advanced semi-automated methods to refine the Ground Truth quality. 

\section*{Impact Statement}
This work introduces \textbf{\AACR\ }, the first multilingual, repository-level context-aware benchmark for Automated Code Review. By comprehensively characterizing the capability boundaries of current Large Language Models, our study identifies critical challenges and charts the course for the next generation of ACR systems.

\paragraph{Paradigm Shift in Automated Code Review}
Our research establishes a rigorous evaluation standard and reveals three fundamental challenges that future ACR methodologies should address to achieve practical utility:
\begin{itemize}
    \item \textbf{Navigating the Precision-Recall Trade-off:} We identify a distinct dichotomy: traditional generative approaches maximize Recall but suffer from frequent hallucinations, whereas Agent-based methods achieve high Precision in defect localization but lack comprehensive coverage. This finding urges the community to move beyond single-metric optimization and focus on hybrid architectures that maintain high precision while expanding defect detection coverage.
    \item \textbf{Towards Adaptive Context Awareness:} We challenge the prevailing assumption that ``more context is always better.'' Our empirical results demonstrate that the impact of retrieved context varies significantly across languages (e.g., C\# vs. Python), where irrelevant context often serves as noise. This necessitates a shift towards \textit{Adaptive Context Awareness}, requiring systems to possess meta-cognitive capabilities to dynamically determine the necessity and granularity of retrieval based on code characteristics and task types.
    \item \textbf{Unifying Local and Global Perspectives:} We observe that while Retrieval-Augmented Generation (RAG) aids global dependency understanding, the induced noise compromises the detection of local defects within Diffs. Conversely, Agents prone to ``context tunneling'' often overlook obvious local errors. Future iterations must develop dynamic attention mechanisms that organically integrate microscopic syntax verification with macroscopic, cross-file risk assessment.
\end{itemize}

\paragraph{From Passive Ingestion to Active Auditing}
This work reframes the evaluation of ``context effectiveness'' in ACR. We argue that high-quality code review depends less on Information Retrieval and more on \textit{Information Utilization} and reasoning robustness. Our findings indicate that without sufficient noise tolerance capability, even relevant code retrieved via BM25 or Embeddings can degrade performance. The superior reasoning of Agents in complex scenarios suggests a paradigm shift from ``Passive Code Ingestion'' (receiving pre-retrieved snippets) to ``Active Code Auditing''. Enabling models to actively explore context, verify hypotheses through multi-turn interactions, and filter noise—mimicking human expert behavior—represents the most promising direction for achieving reliable automated code review.

\bibliography{example_paper}
\bibliographystyle{icml2026}

\newpage
\appendix
\onecolumn
\section{Overview of the Appendix}

The Appendix is organized as follows:

\begin{itemize}
    \item Section~\ref{app:aacr-details} provides detailed information on AACR-Bench, including its data construction process, comparisons with existing benchmarks, and representative data samples.
    \item Section~\ref{app:experiment-detail} describes the experimental setup of AACR-Bench, covering the full evaluation pipeline, tested models, hyperparameter settings, and an analysis of fine-grained results.
    \item Section~\ref{app:case-study} provides case studies based on the AACR-Bench evaluation results
\end{itemize}

\section{AACR-Bench Details} \label{app:aacr-details}

\subsection{Detailed Dataset Curation Process}

\begin{figure}[h]
    \centering
    \includegraphics[width=0.5\linewidth]{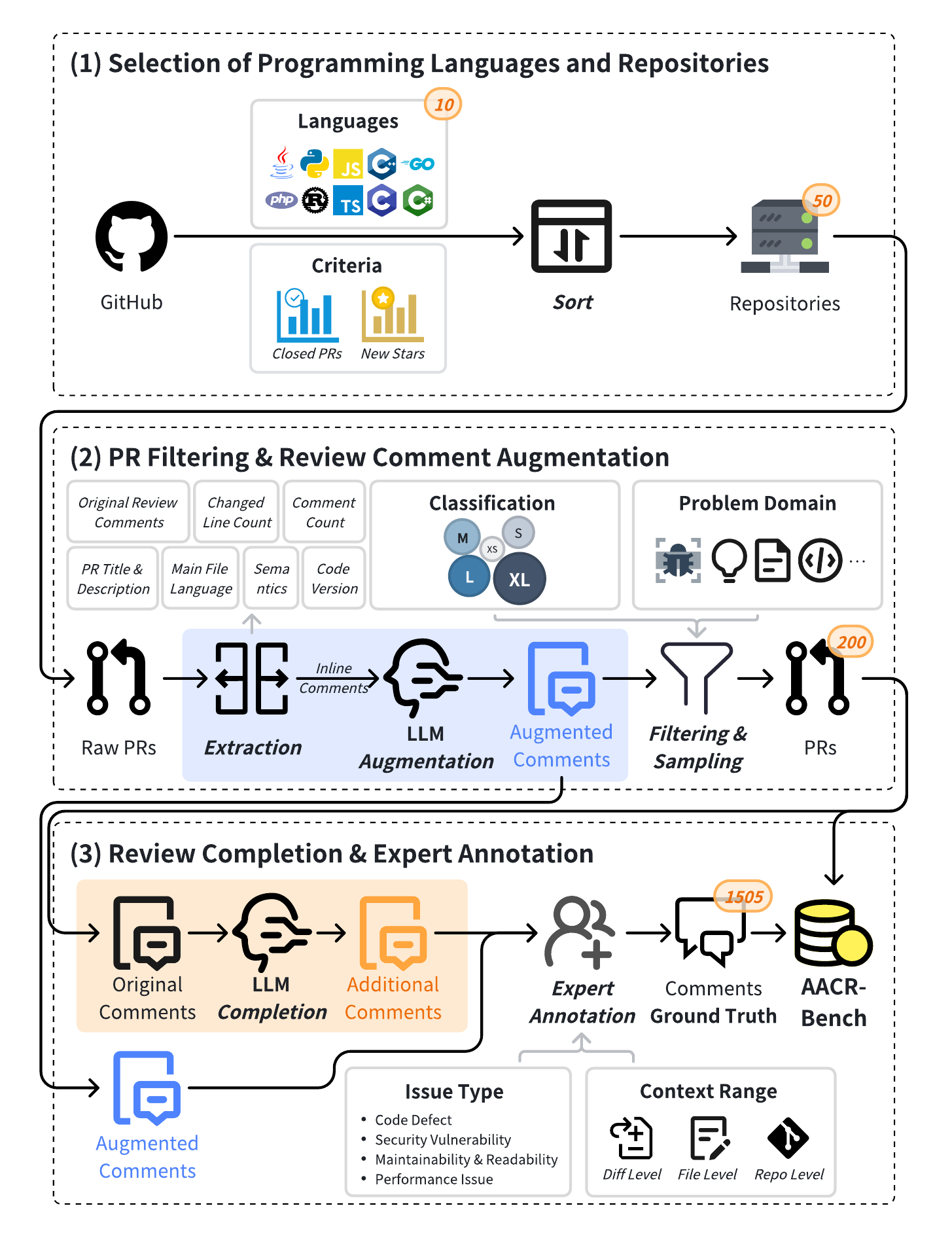}
    \caption{the Construction Process of AACR-Bench}
    \label{fig:construction-process}
\end{figure}

The construction process of AACR-Bench is shown in Figure~\ref{fig:construction-process}.

\paragraph{Selection of Programming Languages and Repositories} To ensure the recency of the evaluation data and the diversity of programming languages, we selected the top ten programming languages based on the StackOverflow Developer Survey 2025, specifically: JavaScript, Python, TypeScript, Java, C\#, C++, C, PHP, Go, and Rust. For each of these languages, we selected repositories where the language serves as the primary programming language to act as data sources for PR information extraction, adhering to the following criteria: First, we defined a candidate set of active GitHub repositories that ranked within the top 2,000 in terms of both new stars and closed PRs between December 1, 2024, and December 1, 2025. Subsequently, we ranked the candidate repositories by the number of new stars and selected the top five for each language. Ultimately, we constructed a collection of 50 high-activity repositories across 10 programming languages (five repositories per language) to serve as the source for extracting PRs and review comments.

\paragraph{PR Filtering and Review Comment Augmentation} Focusing on the 50 selected repositories, we utilized the GitHub API to collect all Pull Requests (PRs) created between December 1, 2024, and December 1, 2025, yielding a total of $12,715$ records. For each PR, we extracted the following key information:

\begin{itemize}
    \item The title and description of the PR;
    \item The number of lines of code changed;
    \item The base commit of the PR;
    \item Comprehensive review comment data, including the target revision, relevant file paths, line ranges, and the specific content of the reviews.
\end{itemize}

Based on the aforementioned raw data, we implemented the following preprocessing steps:

\begin{itemize}
    \item \textbf{Domain Classification:} Analyzed the problem domain of the PR using a Large Language Model (LLM), adhering to the taxonomy defined in SWE-Bench, as shwon in Table~\ref{tab:pr_category};
    \item \textbf{Language Identification:} Analyzed the natural language used in the PR title and description;
    \item \textbf{Revision Selection:} Identified and extracted the revision containing the highest number of review comments, along with all corresponding review data;
    \item \textbf{Size Categorization:} Classified the PR size based on the number of changed lines, following the T-Shirt Size classification method, as shown in Table~\ref{tab:pr_size}.
    \item \textbf{Reviewed Programming Language Identification:} Identified the primary programming language of the reviewed code based on the file extensions of the files targeted by the review comments.
\end{itemize}

\begin{table*}[h]
    \centering
    \caption{Categorization of PR's Problem Domain}
    \label{tab:pr_category}
    \scriptsize
    \begin{tabular}{c p{40em} c}
        \toprule
        \textbf{Category (Abbr.)} & \textbf{Description} & \textbf{PR Count} \\
        \midrule
        Bug Fixes (BF) & Resolving functional errors, crashes, incorrect outputs & 53 \\
        New Feature Additions (NFA) & Adding new functionality or features to the application & 44 \\
        Code Refactoring (CA) & Improving code structure, readability, maintainability without changing external behavior & 28\\
        Documentation Update (DU) & Changes related to code comments or external documentation & 12 \\
        Test Suite / CI Enhancements (TC) & Improving test coverage, test quality, or continuous integration processes & 20 \\
        Performance Optimizations (PO) & Improving application speed, response time, or resource usage efficiency & 19 \\
        Security Patches (SV) & Fixing code defects that could lead to security issues & 13 \\
        Dependency Updates (DE) & Updating third-party library dependencies or ensuring compatibility across different environments & 3 \\
        Code Style, Linting (CLF) & Ensuring code complies with team coding standards and consistency & 8 \\
        \bottomrule
    \end{tabular}
\end{table*}

\begin{table}[]
    \centering
    \scriptsize
    \caption{The T-Shirt Size Categorization of PR Size}
    \label{tab:pr_size}
    \begin{tabular}{c p{12em} c}
        \toprule
        \textbf{Size} & \textbf{Lines Changed} & \textbf{\# PR} \\
        \midrule
        XS & 0 - 9 & 8 \\
        S & 10 -29 & 22 \\
        M & 30 - 99 & 47 \\
        L & 100 - 499 & 89 \\
        XL & 500 - 999 & 34 \\
        XXL & 1000+ & 0 \\
        \bottomrule
    \end{tabular}
\end{table}

For the domain classification task, we employed the Qwen3-235B-A22B-Thinking-2507 model; the specific prompt used is shown in Figure~\ref{fig:prompt-pr-category}. To evaluate the reliability of the classification, we randomly sampled 350 instances from the results for manual verification. The analysis revealed an accuracy of $92.36\%$. At a $95\%$ confidence level, this result has a margin of error of $5.06\%$.

\begin{figure}
    \centering
    \begin{lstlisting}
---Role---

You are a code review assistant tasked with classifying GitHub Pull Requests based on their Title and Description.

---Goal---

Analyze the Title and Description of the provided PR and determine the most appropriate category from the predefined list below. Respond only with the category code (e.g., "BF" or "NFA") without additional text.

The categories are as follows:

- BF: Bug Fix (fixes incorrect behavior, crashes, or errors)
- NFA: New Feature Addition (implements new functionality or capabilities)
- CA: Code Refactoring / Architectural Improvement (improves structure/design without changing behavior)
- DU: Documentation Update (changes to READMEs, comments, or external docs)
- TC: Test Suite / CI Enhancements (adds/fixes tests, CI/CD pipelines, or coverage)
- PO: Performance Optimization (improves speed, memory usage, or resource efficiency)
- SV: Security Patch / Vulnerability Fix (addresses security issues or vulnerabilities)
- DE: Dependency Update / Environment Compatibility (updates libraries, tools, or environment configs)
- CLF: Code Style / Linting / Formatting Fix (enforces style guides, fixes lint errors)

---Examples---

############################

Example 1:

PR Title:
Add Flydigi VADER 4 Pro support

PR Description:
Add a new controller support
## Description
The controller outputs custom HID reports for PS5 controller emulation through vendor software, which also contain motion data unlike the normal HID endpoint.

All inputs checked using `testcontroller`, still not sure how to properly test gyro data and generate mappings for other platforms
---------------
output:
NFA
############################


Example 2:

PR Title:
SDL_BlitSurfaceScaled(): Fix divisions by zero

PR Description:
Related: https://github.com/libsdl-org/SDL/issues/12844
---------------
output:
BF
############################

---Real Data---
############################
PR Title:
{pr_title}

PR Description:
{pr_description}
---------------
output:
    \end{lstlisting}
    \caption{Prompt Used For Classification of PR's Problem Domain}
    \label{fig:prompt-pr-category}
\end{figure}

Building upon the preprocessed data, we established five filtering criteria to automatically select high-quality samples and got 3,328 PR Items:

\begin{itemize}
    \item \textbf{Language Standard:} The title and description of the PR must be written in English.
    \item \textbf{Size Constraint:} The number of changed lines of code must be limited to 1,000. According to Google's code review practices, changes exceeding this threshold are considered too large for effective review.
    \item \textbf{Language Consistency:} The programming language of the primarily modified files must match the main language of the repository.
    \item \textbf{Review Richness:} The PR must contain at least two inline comments, including at least one constructive comment that was accepted (i.e., resulted in code modifications)..
\end{itemize}

We conducted a second round of manual verification on the initially filtered PRs. Focusing on semantic validity, we excluded trivial changes that were detached from the project's business context or lacked actual semantic meaning. This process resulted in a final dataset of 573 PRs.

\begin{figure}
    \centering
    \includegraphics[width=0.8\linewidth]{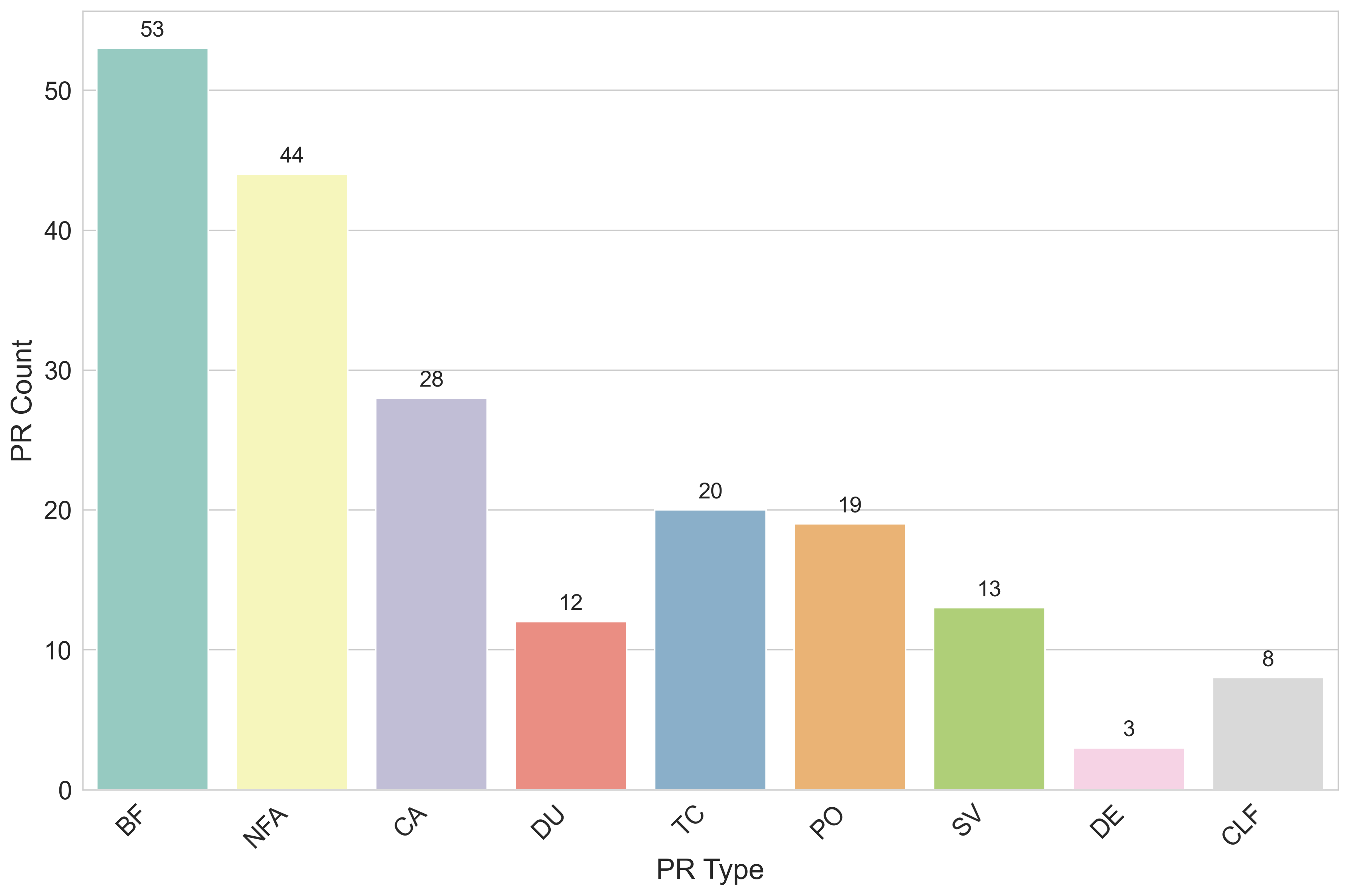}
    \caption{Distribution of PRs Across Problem Domains}
    \label{fig:pr-type-distribution}
\end{figure}

\begin{figure}
    \centering
    \includegraphics[width=0.6\linewidth]{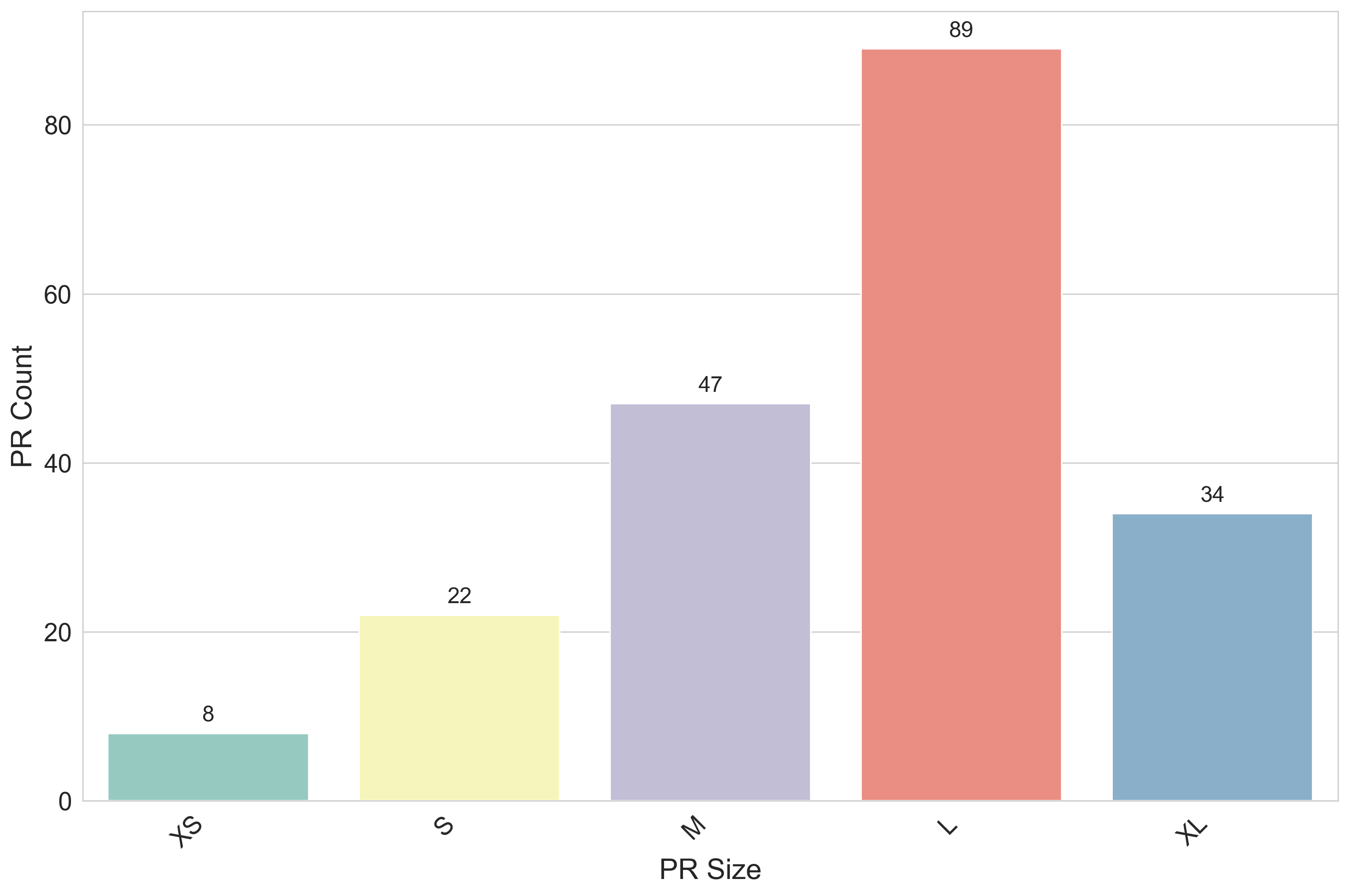}
    \caption{Distribution of PRs Across Different Size}
    \label{fig:pr-size-distribution}
\end{figure}

Finally, to ensure the representativeness and diversity of the benchmark, we applied stratified sampling to the candidate pool based on the source repository, problem domain, and PR size. This process yielded a final dataset consisting of 200 PRs. The distributions of these PRs across problem domains and sizes are shown in Figure~\ref{fig:pr-type-distribution} and Figure~\ref{fig:pr-size-distribution}, respectively.

Given that code review often involves multi-turn dialogues between reviewers and developers to verify the validity of issues, raw comments may suffer from missing context and noise. To address this, we employed an LLM to perform deep semantic analysis on the review threads of the selected revision. Specifically, we extracted confirmed code defects from these multi-round interactions and synthesized them into ``Augmented Review Comments,'' while discarding comments that did not point out substantive issues. The prompt used for this process is shown in Figure~\ref{fig:prompt-cr-augmentation}. We applied this augmentation process to $1,119$ conversation threads and sampled $300$ results for manual evaluation of correctness. An augmentation result was defined as ``correct'' if the model accurately provided review comments identifying confirmed code issues within the thread, or correctly determined that the thread contained no code issues. The evaluation results indicate an accuracy of 95\%, with a 95\% confidence level and a 4.74\% margin of error.

\begin{figure}
    \centering
    \begin{lstlisting}
## Code Review Summary Prompt

You are a senior code reviewer tasked with summarizing code review discussions. Given a series of conversations and their corresponding diff hunks, your goal is to extract the most concise and actionable review feedback.

### Instructions:

1. **Analyze the provided materials:**
   - Review conversations between developers
   - Examine the associated code diff hunks
   - Identify the core issues being discussed

2. **Summarize with precision:**
   - Extract the main concern or issue in 1-2 sentences
   - Include only essential context needed to understand the problem
   - Focus on technical substance, not conversational details
   - Be direct and specific about what needs attention

3. **Format your response:**
   - Start with the core issue
   - Follow with minimal necessary context
   - Use clear, technical language
   - Avoid redundancy or filler words

4. **Special cases:**
   - If the discussion concludes that no issues exist, simply respond: "No code issues identified."
   - If multiple issues are discussed, list them as bullet points, each with its own concise summary

### Example Output Format:
[Brief background information if necessary] [One-sentence description of the problem]

Or for no issues:
"No code issues identified."

### Your Task:
Review the following conversation and diff hunk, then provide your summary following the above guidelines.

Diff Hunk:
{diff_hunk}

Conversation:
{review_comments}
    \end{lstlisting}
    \caption{Prompt Used For Review Comments Augmentation}
    \label{fig:prompt-cr-augmentation}
\end{figure}

\begin{table*}[]
    \centering
    \caption{Issue Categories revealed in Review Comments}
    \scriptsize
    \begin{tabular}{c p{40em} c}
        \toprule
        \textbf{Issue Category} & \textbf{Description} & \textbf{Comment Count} \\
        \midrule
        Security Vulnerability & Security weaknesses within the code that could lead to data breaches, unauthorized access, or susceptibility to other malicious attacks & 53 \\
        Code Defect & Logical errors or implementation flaws that may cause runtime crashes, produce incorrect results, or result in unexpected system behaviors & 709 \\
        Maintainability \& Readability & Issues related to coding style or structural design that diminish code readability and hinder future comprehension and maintenance efforts & 626 \\
        Performance Issue & Inefficiencies in algorithms or improper resource management that result in non-functional bottlenecks, such as high latency, insufficient throughput, or excessive resource consumption & 117 \\
        \bottomrule
    \end{tabular}
    \label{tab:comment_issue_category}
\end{table*}

\paragraph{Review Completion and Expert Annotation} Constrained by the cognitive boundaries and capacity limits of human reviewers, relying solely on manual efforts often fails to uncover all potential code defects. To address the resulting issue of ``Incomplete Issue Annotation,'' we employed LLM generation techniques to comprehensively augment the review comments. To mitigate single-model bias and ensure diversity, we constructed a generation matrix comprising six mainstream open-source and proprietary models. These models generated review comments in parallel through two heterogeneous frameworks: an Internal Review System and an Open-Source Agent System (Claude Code). All generated comments underwent semantic de-duplication before being merged with the previously augmented human review comments, forming a candidate set for verification. This semantic de-duplication was performed using Qwen3-235B-A22B-Thinking-2507. Specifically, all review comments were first grouped based on their repository, Pull Request (PR), file path, and the specific Diff Hunk addressed. Within each group, comments underwent pairwise comparison via the LLM, and duplicates were removed based on the comparison results. The prompt used for this comparison is shown in Figure~\ref{fig:prompt-repeat}. We use the election of results running 5 times of judgement.

\begin{figure}
    \centering
    \begin{lstlisting}
-Role-

You are an expert code reviewer assistant specialized in analyzing and comparing code review comments.

-Task-

Determine whether two given review comments express the same concern or suggestion regarding the privided diff hunk. Ignore differences in wording, tone, or formatting-focus solely on semantic equivalence of the underlying issue. If the core intent and technical substance are identical, answer "yes", otherwise, answer "no".

-Diff Hunk-

{diff_hunk}

-Review Comments-

Review Comment 1:
{comment1}

Review Comment 2:
{comment2}

-Task-

Determine whether the two review comments given above express the same concern or suggestion regarding the provided diff hunk. Ignore differences in wording, tone, or formatting-focus solely on semantic equivalence of the underlying issue. If the core intent and technical substance are identical, answer "yes"; other wise, answer "no".

Your answer:
    \end{lstlisting}
    \caption{Prompt Used to detect Repeated Comments}
    \label{fig:prompt-repeat}
\end{figure}

Subsequently, we submitted this set to over 80 senior software engineers with more than two years of experience for rigorous Human Annotation. The annotation process covered three core dimensions: verifying the correctness of the review comments, categorizing issue types according to Table~\ref{tab:comment_issue_category}, and defining the scope of context dependency. The annotation workforce consisted of a Core Expert Team of 6 members and a General Annotation Pool comprising the remaining participants. The annotation process was divided into three rounds: The first two rounds were conducted by the General Annotation Pool using a double-blind mechanism, where each comment was independently annotated by two different individuals, and task allocation was strictly matched to the annotators' programming language expertise. The third round was conducted by the Core Expert Team, which was responsible for discussing and adjudicating conflicting results from the first two rounds and determining the final annotations. Through this ``human-machine collaborative'' multi-round annotation workflow, we ensured that the final evaluation benchmark possesses both high coverage and high accuracy.

To evaluate the complementarity of the chosen LLMs, we analyzed the overlap of generated review comments along with augmented review comments. Table~\ref{tab:comments-models-dist} illustrates the distribution of comments based on the number of models that successfully detected them in the annotated ground truth.

\begin{table}[]
    \centering
    \caption{Distribution of Review Comments by Model Detection Frequency}
    \label{tab:comments-models-dist}
    \begin{tabular}{cc}
        \toprule
        \textbf{Number of Detecting Models} & \textbf{Count} \\
        \midrule
        0(Augmented only) & 360 \\
        1 & 1027 \\
        2 & 107 \\
        3 & 11 \\
        \bottomrule
    \end{tabular}
\end{table}

Interestingly, while only $11$ of the comments achieved consensus among 3 models (detected by three models), a significant portion was identified by only a single model. This low overlap suggests the necessity to include multi-model results in review comment generation process for better issue coverage.



\subsection{Examples of Ground Truth}

In this section, we present examples of review comments that were verified as valid through the final annotation process, thereby serving as the Ground Truth for AACR-Bench. For each example, we provide the repository name, the Pull Request (PR) ID, the target file, the specific Diff Hunk addressed, the content of the review comment, and a corresponding analysis.

\begin{figure}
    \centering
    \includegraphics[width=\linewidth]{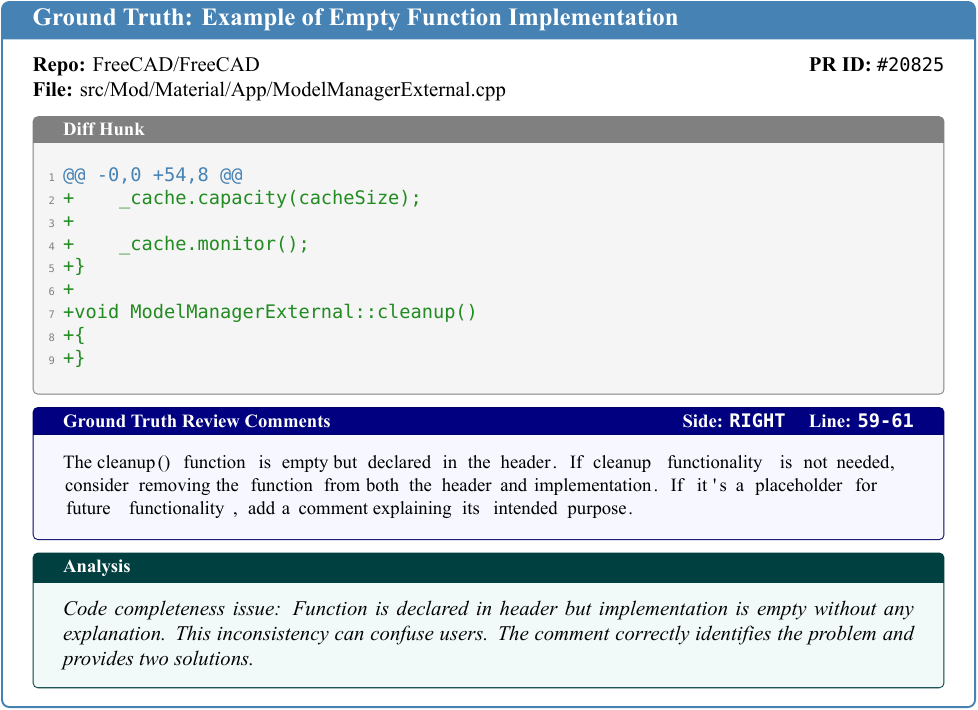}
    \caption{An Example of Gound Truth Reveiw Comment}
    \label{fig:ground-1}
\end{figure}

\begin{figure}
    \centering
    \includegraphics[width=\linewidth]{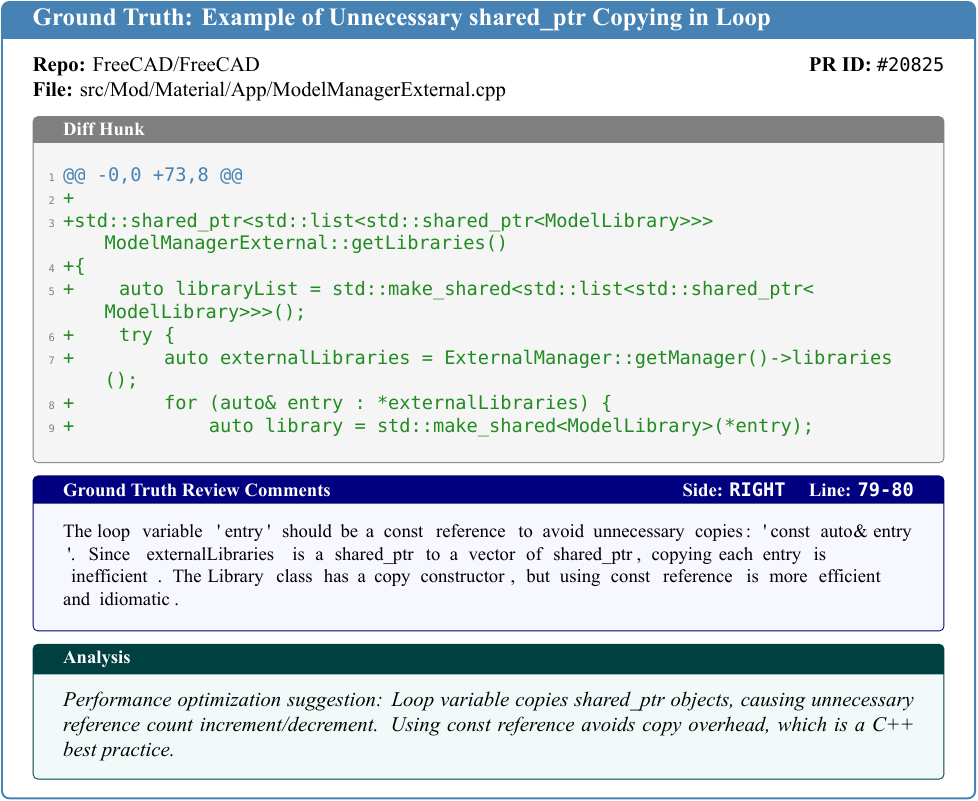}
    \caption{An Example of Gound Truth Reveiw Comment}
    \label{fig:ground-2}
\end{figure}

\begin{figure}
    \centering
    \includegraphics[width=\linewidth]{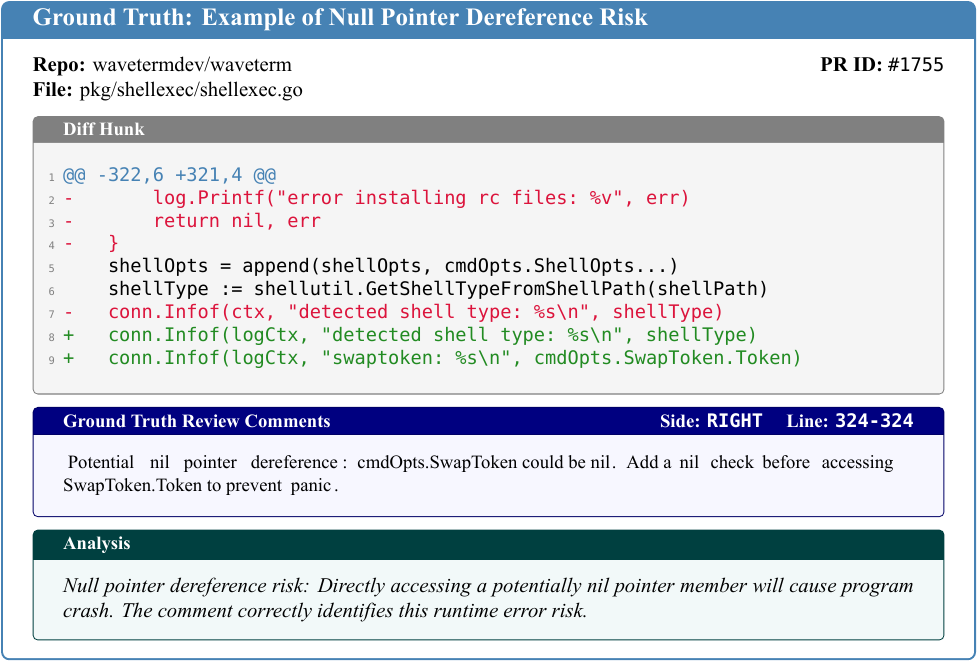}
    \caption{An Example of Gound Truth Reveiw Comment}
    \label{fig:ground-3}
\end{figure}

\begin{figure}
    \centering
    \includegraphics[width=\linewidth]{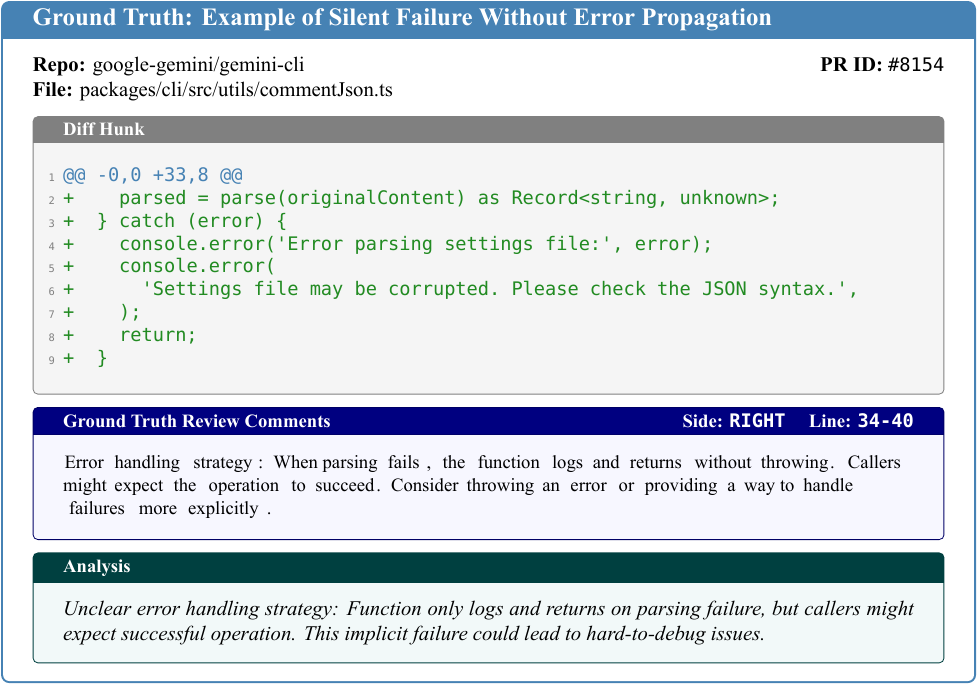}
    \caption{An Example of Gound Truth Reveiw Comment}
    \label{fig:ground-4}
\end{figure}

\begin{figure}
    \centering
    \includegraphics[width=\linewidth]{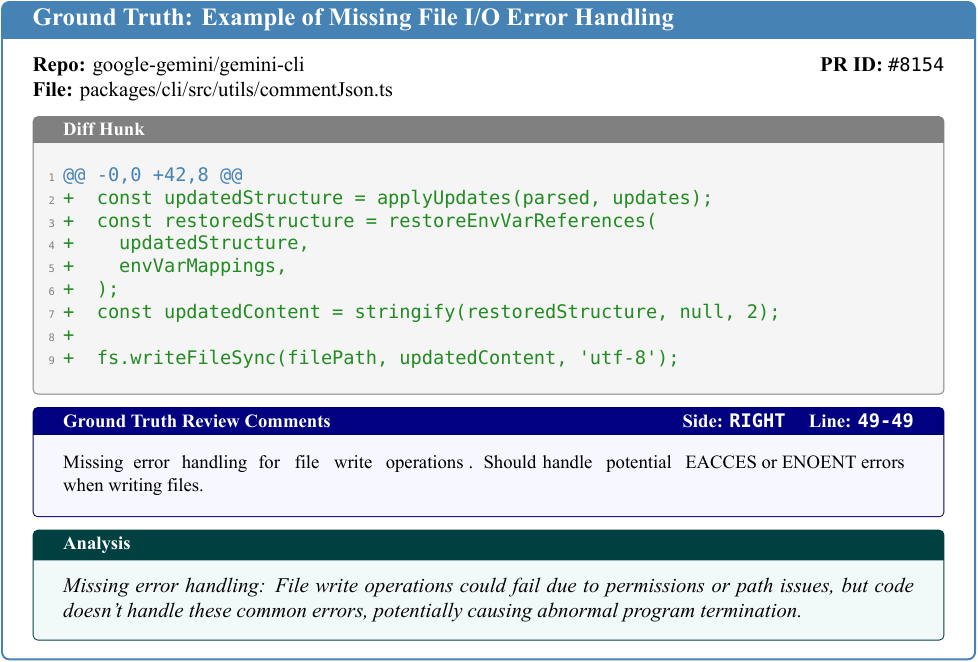}
    \caption{An Example of Gound Truth Reveiw Comment}
    \label{fig:ground-5}
\end{figure}

\begin{figure}
    \centering
    \includegraphics[width=\linewidth]{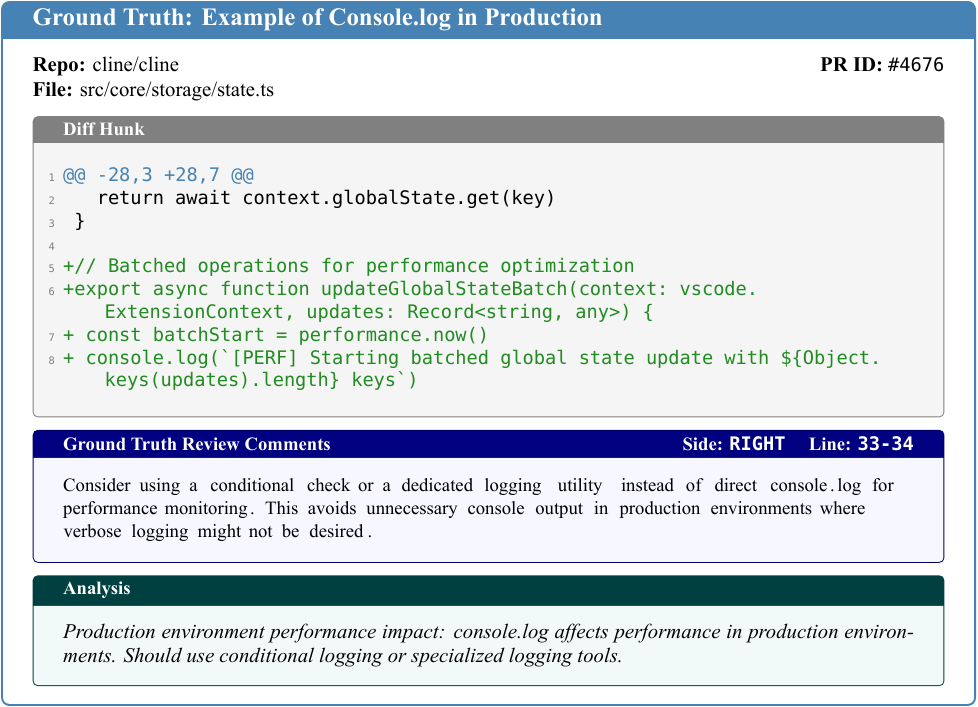}
    \caption{An Example of Gound Truth Reveiw Comment}
    \label{fig:ground-6}
\end{figure}

\section{Detail of Experiments} \label{app:experiment-detail}

\subsection{Detailed Benchmark Process}

The evaluation process of AACR-Bench aligns with the Diff-oriented paradigm of modern code review. Within the benchmark, each Pull Request (PR) serves as an evaluation instance, where the collection of review comments associated with that PR constitutes the Ground Truth. For each PR, the Automated Code Review (ACR) method under evaluation is required to iterate through the Diff Hunks within the code changes and generate review comments for each hunk. The code review capability of the ACR method is then measured by comparing the generated comments against the Ground Truth to calculate Precision, Recall, and F1-score. Prompt for review comment match is the same as~\ref{fig:prompt-repeat} and the model used is Qwen3-235B-A22B-Instruct-2507. A generated review comment is considered Line Correct if its line range overlaps with that of a review comment in the ground truth. It is considered Semantically Correct if it overlaps with the ground truth in terms of line numbers and the content is identical. All the models are run with $Temperature = 0.7$, $Top_p=0.95$, $Top_k=20$.

\paragraph{Evaluation of non-Agent Methods} The evaluation of non-Agent approaches follows a standardized workflow. For each PR in the benchmark, we first clone the repository locally and ensure that both the Base version and the Target version (i.e., the version under review) are synchronized. We then verify and extract all diff hunks between the two versions using GitPython. The ACR method under evaluation acts as a scanner, iterating through each diff hunk to generate review comments. For methods incorporating context retrieval, we build an index based on the code in the Target version to facilitate context recall. During the review of each Diff Hunk, we invariably provide the PR Title and PR Description as repository-level context. Additionally, repository code context is provided optionally, subject to the specific experimental design. The prompts used for evaluation are illustrated in Figure~\ref{fig:prompt-cr-nocode} and Figure~\ref{fig:prompt-cr-code}. Specifically, for each review comment, the model is required to output three components:

\begin{itemize}
    \item The problematic diff fragment: Used for precise localization of the comment;
    \item The comment side: Specifying whether the comment applies to the 'left' (deleted lines in the old version) or the 'right' (added lines in the new version);
    \item The content of the review comment.
\end{itemize}

\begin{figure}
    \centering
    \begin{lstlisting}
You are an expert code reviewer. Your task is to review the following code changes and provide constructive feedback.

## Pull Request Information

**Title:** {{ pr_title }}

**Description:** {{ pr_description }}

## Code Changes to Review

```diff
{{ diff_hunk }}
```

## Instructions

Please review the code changes above and provide detailed feedback. For each issue you identify:

1. Use `<diff>...</diff>` tags to wrap **only the minimal code snippet** that contains the specific issue you're commenting on. Do NOT include the entire diff_hunk - extract only the relevant lines that demonstrate the problem.
2. Use `<side>left</side>` or `<side>right</side>` to indicate whether your comment applies to the old code (left) or new code (right)
3. Use `<note>...</note>` tags to wrap your review comment
4. Use `<notesplit />` to separate different review comments
5. Use `<end />` to end your response

**Important**: The `<diff>` section should contain the smallest possible code fragment that clearly shows the issue. If multiple separate issues exist in different parts of the diff_hunk, create separate review comments for each with their own minimal `<diff>` snippets.

Focus on:
- Code defects
- Performance issues
- Security vulnerability
- Maintainability and readability

## Example Format

<diff>
+ def example_function():
+     pass
</diff>
<side>right</side>
<note>
This function lacks documentation. Please add a docstring explaining its purpose, parameters, and return value.
</note>
<notesplit />
<end />

## Output Requirements

- If you find issues with the code changes, provide your review comments following the format above
- If you believe the code changes are acceptable and have no issues to report, simply output `<end />`
- Always end your response with `<end />` after all review comments (or immediately if no issues found)

Now provide your review:
    \end{lstlisting}
    \caption{Prompt for ACR Tasks of non-Agent Methods with no Code Context}
    \label{fig:prompt-cr-nocode}
\end{figure}

\begin{figure}
    \centering
    \begin{lstlisting}
You are an expert code reviewer. Your task is to review the following code changes and provide constructive feedback.

## Pull Request Information

**Title:** {{ pr_title }}

**Description:** {{ pr_description }}

## Code Changes to Review

```diff
{{ diff_hunk }}
```

## Relevant Code Context

{{ contexts }}

## Instructions

Please review the code changes above and provide detailed feedback. For each issue you identify:

1. Use `<diff>...</diff>` tags to wrap **only the minimal code snippet** that contains the specific issue you're commenting on. Do NOT include the entire diff_hunk - extract only the relevant lines that demonstrate the problem.
2. Use `<side>left</side>` or `<side>right</side>` to indicate whether your comment applies to the old code (left) or new code (right)
3. Use `<note>...</note>` tags to wrap your review comment
4. Use `<notesplit />` to separate different review comments
5. Use `<end />` to end your response

**Important**: The `<diff>` section should contain the smallest possible code fragment that clearly shows the issue. If multiple separate issues exist in different parts of the diff_hunk, create separate review comments for each with their own minimal `<diff>` snippets.

Focus on:
- Code defects
- Performance issues
- Security vulnerability
- Maintainability and readability

## Example Format

<diff>
+ def example_function():
+     pass
</diff>
<side>right</side>
<note>
This function lacks documentation. Please add a docstring explaining its purpose, parameters, and return value.
</note>
<notesplit />
<end />

## Output Requirements

- If you find issues with the code changes, provide your review comments following the format above
- If you believe the code changes are acceptable and have no issues to report, simply output `<end />`
- Always end your response with `<end />` after all review comments (or immediately if no issues found)

Now provide your review:
    \end{lstlisting}
    \caption{Prompt for ACR Tasks of non-Agent Methods with Code Context}
    \label{fig:prompt-cr-code}
\end{figure}

\paragraph{Evaluation of Agent Methods} We selected Claude Code as the framework for evaluating Agent-based approaches. Leveraging its autonomous tool-use capabilities, Claude Code can independently invoke Git commands to extract Diff Hunks between the Base and Target versions and perform fully autonomous context retrieval. Consequently, the evaluation setup is streamlined: we typically clone the repository and checkout the corresponding PR version locally, eliminating the need for pre-built code indexing. The definition of the Code Review Agent utilized in this evaluation is illustrated in Figure~\ref{fig:agent-def}, while the specific prompt used to trigger the review process is presented in Figure~\ref{fig:agent-prompt}.

\begin{figure}
    \centering
    \begin{lstlisting}
---
name: code-reviewer
description: Code review and post inline comments if issues are found
tools: Read,Write,Bash
model: inherit
---

Provide a code review.

To do this, follow these steps precisely:

1. Identify possible issue in MR
   We do NOT want:

- Subjective concerns or "suggestions"
- Style preferences not explicitly required by CLAUDE.md
- Potential issues that "might" be problems
- Anything requiring interpretation or judgment calls

If you are not certain an issue is real, do not flag it. False positives erode trust and waste reviewer time.

In addition to the above, each subagent should be told the PR title and description. This will help provide context
regarding the author intent.

2. For each issue found, launch parallel subagents to validate the issue. These subagents should get the PR title and
   description along with a description of the issue.
   The agent job is to review the issue to validate that the stated issue is truly an issue with high confidence. For
   example, if an issue such as "variable is not defined" was flagged, the subagent job would be to validate that is
   actually true in the code.

3. If issues were found, post inline comments for **EACH ISSUE** using below style:
    ```
    <<path>[file_path]</path>
    <side>left</side> or <side>right</side>
    <from>[begin_line_number]</from>
    <to>[end_line_number]</to>
    <note>[detailed_review_comment]</note>
    <notesplit />
    ```
   **Suggestions must be COMPLETE.** If a fix requires additional changes elsewhere (e.g., renaming a variable requires
   updating all usages).

For larger fixes (6+ lines, structural changes, or changes spanning multiple locations), do NOT use suggestion blocks.
Instead:

1. Describe what the issue is
2. Explain the suggested fix at a high level
3. Include a copyable prompt for Claude Code that the user can use to fix the issue, formatted as:
   ```
   Fix [file:line]: [brief description of issue and suggested fix]
   ```
   **IMPORTANT: Only post ONE comment per unique issue. Do not post duplicate comments.**

Use this list when evaluating issues (these are false positives, do NOT flag):

- Pre-existing issues
- Something that appears to be a bug but is actually correct
- Pedantic nitpicks that a senior engineer would not flag
- Issues that a linter will catch (do not run the linter to verify)
- General code quality concerns (e.g., lack of test coverage, general security issues) unless explicitly required in
  CLAUDE.md
- Issues mentioned in CLAUDE.md but explicitly silenced in the code (e.g., via a lint ignore comment)

Notes:

- Create a todo list before starting.
- You must cite and link each issue in inline comments (e.g., if referring to a CLAUDE.md, include a link to it).
- FOR EACH ISSUE YOU FIND, POST A COMMENT FOR EACH ISSUE.

**REMEMBER TO POST A COMMENT IF ANY ISSUES WERE FOUND**
    \end{lstlisting}
    \caption{Agent Definition of Claude Code}
    \label{fig:agent-def}
\end{figure}

\begin{figure}
    \centering
    \begin{lstlisting}
Please conduct a code review for the changes from {base_commit}...{target_commit}, with PR title: {pr_title}
PR description: {pr_description}
**write the review comments to comments.txt(if the file exists, clear it first)**,
following the format:
<path>[file_path]</path>
<side>left</side> or <side>right</side>
<from>[begin_line_number]</from>
<to>[end_line_number]</to>
<note>[detailed_review_comment]</note>
<notesplit />
**If you think there are no issues with the code, there's no need to create the file or write comments**.
If possible, use code-reviewer first.
    \end{lstlisting}
    \caption{Agent Prompt of Claude Code}
    \label{fig:agent-prompt}
\end{figure}

To evaluate the model's accuracy in detecting different categories of issues, we classified the review comments generated by the ACR methods. This classification was performed using Qwen3-235B-A22B-Instruct-2507, utilizing the prompt illustrated in Figure~\ref{fig:prompt-issue-classification}. We validated the efficacy of this prompt on the benchmark's Ground Truth dataset, achieving an accuracy of $97\%$.

\begin{figure}
    \centering
    \begin{lstlisting}
---Role---

You are a senior software engineer specializing in code quality, security, and performance. You have extensive experience reviewing code changes and identifying underlying issues based on contextual feedback.

---Task---

Given the code diff hunk and review comment below, classify the reported issue into exactly one of the following categories using only the category name as your output-no explanations, punctuation, or extra text.

Category definitions:

- **Security Vulnerability:** A point in your code that's open to attack.
- **Code Defect:** A coding mistake that can lead to an error or unexpected behavior at runtime
- **Maintainability and Readability:** A maintainability issue that makes your code confusing and difficult to maintain.
- **Performance:** A non-functional defect caused by inefficient algorithms, improper resource usage (e.g., memory, CPU, I/O), poor concurrency handling, or architectural flaws, resulting in excessive response times, insufficient throughput, high resource consumption, or limited scalability.

---Diff Hunk & Review Comment---

Diff Hunk:
{diff_hunk}

Review Comment:
{review_comment}
==============================
Output:
    \end{lstlisting}
    \caption{Prompt for Classification of Issues Found By ACR methods}
    \label{fig:prompt-issue-classification}
\end{figure}

\subsection{Additional Statistics}

\begin{table}[hp]
\scriptsize
    \centering
    \caption{Detailed Performance Comparison (\%) of Models Across Different Types}
    \label{tab:model_performance_detailed}
    \resizebox{\textwidth}{!}{%
        \begin{tabular}{llccccccc}
            \toprule
            \textbf{Model} & \textbf{Type} & \textbf{Avg. Comments} & \textbf{Line} & \textbf{Semantic} & \textbf{Line} & \textbf{Semantic} & \textbf{Line} & \textbf{Semantic} \\
             &  & \textbf{(per Patch)} & \textbf{Recall} & \textbf{Recall} & \textbf{Precision} & \textbf{Precision} & \textbf{F1} & \textbf{F1} \\
            \midrule
            Claude-4.5-Sonnet & Agent & 0.08 & \cc{18}{13.16} & \cc{21}{10.10} & \cc{100}{52.00} & \cc{100}{39.90} & \cc{70}{21.00} & \cc{100}{16.12}\\
            Deepseek-V3.2 & Agent & 0.15 & \cc{19}{14.15} & \cc{10}{4.78} & \cc{63}{32.60} & \cc{28}{11.00} & \cc{66}{19.74} & \cc{41}{6.67}\\
            GLM-4.7 & Agent & 0.14 & \cc{13}{9.63} & \cc{10}{4.72} & \cc{45}{23.40} & \cc{29}{11.50} & \cc{46}{13.65} & \cc{41}{6.69}\\
            GPT-5.2 & Agent & 0.11 & \cc{7}{5.51} & \cc{6}{2.99} & \cc{35}{18.20} & \cc{25}{9.90} & \cc{28}{8.46} & \cc{28}{4.59}\\
            Qwen-480B-Coder & Agent & 0.10 & \cc{14}{10.63} & \cc{9}{4.39} & \cc{72}{37.20} & \cc{38}{15.30} & \cc{55}{16.54} & \cc{42}{6.82}\\
            \midrule
            Claude-4.5-Sonnet & No context & 1.72 & \cc{98}{73.29} & \cc{91}{42.86} & \cc{29}{15.00} & \cc{22}{8.70} & \cc{83}{24.90} & \cc{90}{14.46} \\
            Deepseek-V3.2 & No context & 2.29 & \cc{98}{72.76} & \cc{77}{36.54} & \cc{22}{11.20} & \cc{14}{5.60} & \cc{65}{19.41} & \cc{60}{9.71}\\
            GLM-4.7 & No context & 0.85 & \cc{67}{50.23} & \cc{58}{27.57} & \cc{40}{20.60} & \cc{28}{11.30} & \cc{97}{29.22} & \cc{99}{16.03} \\
            GPT-5.2 & No context & 2.35 & \cc{99}{73.89} & \cc{100}{47.11} & \cc{21}{11.00} & \cc{18}{7.00} & \cc{64}{19.15} & \cc{76}{12.19} \\
            Qwen-480B-Coder & No context & 1.02 & \cc{78}{58.34} & \cc{58}{27.44} & \cc{39}{20.10} & \cc{24}{9.40} & \cc{100}{29.90} & \cc{87}{14.00} \\
            \midrule
            Claude-4.5-Sonnet & BM25 & 2.17 & \cc{97}{72.56} & \cc{76}{35.75} & \cc{23}{11.80} & \cc{15}{5.80} & \cc{68}{20.30} & \cc{62}{9.98}\\
            Deepseek-V3.2 & BM25 & 0.89 & \cc{69}{51.69} & \cc{58}{27.38} & \cc{39}{20.50} & \cc{27}{10.90} & \cc{98}{29.36} & \cc{97}{15.59} \\
            GLM-4.7 & BM25 & 0.91 & \cc{72}{53.62} & \cc{56}{26.25} & \cc{40}{20.80} & \cc{26}{10.20} & \cc{100}{29.97} & \cc{91}{14.69} \\
            GPT-5.2 & BM25 & 1.75 & \cc{100}{74.55} & \cc{92}{43.59} & \cc{29}{15.00} & \cc{22}{8.80} & \cc{83}{24.97} & \cc{91}{14.64} \\
            Qwen-480B-Coder & BM25 & 2.39 & \cc{97}{72.49} & \cc{97}{45.85} & \cc{21}{10.70} & \cc{17}{6.70} & \cc{62}{18.65} & \cc{73}{11.69} \\
                        \midrule
            Claude-4.5-Sonnet & Embedding & 1.89 & \cc{98}{73.16} & \cc{91}{42.86} & \cc{26}{13.60} & \cc{20}{8.00} & \cc{77}{22.94} & \cc{84}{13.48} \\
            Deepseek-V3.2 & Embedding & 2.52 & \cc{97}{72.56} & \cc{77}{36.35} & \cc{19}{10.10} & \cc{13}{5.10} & \cc{59}{17.73} & \cc{55}{8.94}\\
            GLM-4.7 & Embedding & 0.87 & \cc{67}{49.63} & \cc{57}{26.98} & \cc{39}{20.20} & \cc{28}{11.00} & \cc{96}{28.71} & \cc{97}{15.63} \\
            GPT-5.2 & Embedding & 2.47 & \cc{97}{72.49} & \cc{100}{47.24} & \cc{20}{10.30} & \cc{17}{6.70} & \cc{60}{18.04} & \cc{73}{11.74} \\
            Qwen-480B-Coder & Embedding & 0.83 & \cc{66}{49.50} & \cc{51}{24.25} & \cc{40}{20.90} & \cc{26}{10.20} & \cc{98}{29.39} & \cc{89}{14.36} \\
            \bottomrule
        \end{tabular}%
    }
\end{table}

\begin{table}[hp]
\centering
\caption{Performance Comparison (\%) of Different Models and Methods across Four Issue Categories}
\label{tab:model_comparison}
\resizebox{\textwidth}{!}{%
\begin{tabular}{llcccccccccccc}
\toprule
\multirow{2}{*}{\textbf{Model}} & \multirow{2}{*}{\textbf{Method}} & \multicolumn{3}{c}{\textbf{Security Vuln.}} & \multicolumn{3}{c}{\textbf{Code Defect}} & \multicolumn{3}{c}{\textbf{Performance}} & \multicolumn{3}{c}{\textbf{Maint. \& Read.}} \\
\cmidrule(lr){3-5} \cmidrule(lr){6-8} \cmidrule(lr){9-11} \cmidrule(lr){12-14}
 &  & \textbf{Rec} & \textbf{Prec} & \textbf{F1} & \textbf{Rec} & \textbf{Prec} & \textbf{F1} & \textbf{Rec} & \textbf{Prec} & \textbf{F1} & \textbf{Rec} & \textbf{Prec} & \textbf{F1} \\
\midrule
Claude-4.5-Sonnet & Agent & \cc{20}{11.76} & \cc{100}{17.39} & \cc{92}{14.04} & \cc{33}{17.07} & \cc{100}{44.02} & \cc{100}{24.60} & \cc{17}{9.28} & \cc{100}{36.00} & \cc{57}{14.75} & \cc{27}{11.30} & \cc{100}{37.60} & \cc{100}{17.38} \\
Deepseek-V3.2 & Agent & \cc{4}{2.08} & \cc{12}{2.17} & \cc{14}{2.13} & \cc{13}{6.71} & \cc{31}{13.65} & \cc{37}{8.99} & \cc{5}{2.65} & \cc{16}{5.88} & \cc{14}{3.66} & \cc{9}{3.68} & \cc{26}{9.69} & \cc{31}{5.33} \\
GLM-4.7 & Agent & \cc{8}{4.88} & \cc{23}{4.08} & \cc{29}{4.44} & \cc{13}{6.72} & \cc{32}{14.29} & \cc{37}{9.14} & \cc{16}{8.42} & \cc{56}{20.00} & \cc{46}{11.85} & \cc{11}{4.67} & \cc{22}{8.43} & \cc{35}{6.01} \\
GPT-5.2 & Agent & \cc{7}{4.00} & \cc{41}{7.14} & \cc{34}{5.13} & \cc{7}{3.40} & \cc{19}{8.36} & \cc{20}{4.84} & \cc{5}{2.59} & \cc{49}{17.65} & \cc{17}{4.51} & \cc{6}{2.56} & \cc{33}{12.31} & \cc{24}{4.24} \\
Qwen-480B-Coder & Agent & \cc{7}{4.17} & \cc{77}{13.33} & \cc{42}{6.35} & \cc{13}{6.46} & \cc{41}{18.22} & \cc{39}{9.53} & \cc{7}{4.00} & \cc{46}{16.67} & \cc{25}{6.45} & \cc{8}{3.49} & \cc{30}{11.31} & \cc{31}{5.33} \\
\midrule
Claude-4.5-Sonnet & No context & \cc{77}{45.28} & \cc{28}{4.81} & \cc{57}{8.70} & \cc{90}{46.40} & \cc{36}{15.63} & \cc{95}{23.38} & \cc{100}{53.85} & \cc{39}{14.13} & \cc{86}{22.38} & \cc{87}{36.58} & \cc{14}{5.30} & \cc{53}{9.26} \\
Deepseek-V3.2 & No context & \cc{48}{28.30} & \cc{20}{3.45} & \cc{40}{6.15} & \cc{77}{39.63} & \cc{24}{10.35} & \cc{67}{16.42} & \cc{81}{43.59} & \cc{31}{11.02} & \cc{68}{17.59} & \cc{77}{32.43} & \cc{9}{3.28} & \cc{34}{5.96} \\
GLM-4.7 & No context & \cc{65}{37.74} & \cc{55}{9.57} & \cc{100}{15.27} & \cc{56}{28.77} & \cc{40}{17.82} & \cc{89}{22.01} & \cc{65}{35.04} & \cc{58}{20.71} & \cc{100}{26.03} & \cc{57}{23.96} & \cc{19}{7.08} & \cc{63}{10.93} \\
GPT-5.2 & No context & \cc{97}{56.60} & \cc{17}{2.92} & \cc{36}{5.56} & \cc{98}{50.78} & \cc{32}{13.92} & \cc{89}{21.85} & \cc{89}{47.86} & \cc{27}{9.82} & \cc{63}{16.30} & \cc{100}{42.01} & \cc{12}{4.46} & \cc{46}{8.07} \\
Qwen-480B-Coder & No context & \cc{45}{26.42} & \cc{45}{7.87} & \cc{79}{12.12} & \cc{58}{29.90} & \cc{39}{16.99} & \cc{88}{21.67} & \cc{57}{30.77} & \cc{39}{14.17} & \cc{75}{19.41} & \cc{57}{24.12} & \cc{15}{5.61} & \cc{52}{9.10} \\
\midrule
Claude-4.5-Sonnet & BM25 & \cc{81}{47.17} & \cc{28}{4.84} & \cc{57}{8.77} & \cc{94}{48.38} & \cc{36}{15.92} & \cc{97}{23.95} & \cc{90}{48.72} & \cc{34}{12.28} & \cc{75}{19.62} & \cc{88}{36.90} & \cc{14}{5.32} & \cc{54}{9.31} \\
Deepseek-V3.2 & BM25 & \cc{58}{33.96} & \cc{23}{3.97} & \cc{47}{7.11} & \cc{77}{39.49} & \cc{25}{10.80} & \cc{69}{16.96} & \cc{68}{36.75} & \cc{26}{9.39} & \cc{57}{14.96} & \cc{75}{31.47} & \cc{9}{3.41} & \cc{35}{6.15} \\
GLM-4.7 & BM25 & \cc{52}{30.19} & \cc{41}{7.08} & \cc{75}{11.47} & \cc{57}{29.20} & \cc{41}{18.00} & \cc{91}{22.27} & \cc{57}{30.77} & \cc{53}{18.95} & \cc{90}{23.45} & \cc{58}{24.44} & \cc{18}{6.88} & \cc{62}{10.74} \\
GPT-5.2 & BM25 & \cc{97}{56.60} & \cc{17}{2.97} & \cc{37}{5.64} & \cc{95}{49.08} & \cc{30}{13.27} & \cc{85}{20.89} & \cc{87}{47.01} & \cc{27}{9.79} & \cc{62}{16.20} & \cc{98}{41.05} & \cc{11}{4.26} & \cc{44}{7.71} \\
Qwen-480B-Coder & BM25 & \cc{35}{20.75} & \cc{40}{6.92} & \cc{68}{10.38} & \cc{55}{28.21} & \cc{43}{18.94} & \cc{92}{22.66} & \cc{46}{24.79} & \cc{38}{13.62} & \cc{68}{17.58} & \cc{59}{24.76} & \cc{17}{6.31} & \cc{58}{10.06} \\
\midrule
Claude-4.5-Sonnet & Embedding & \cc{77}{45.28} & \cc{26}{4.45} & \cc{53}{8.11} & \cc{91}{46.83} & \cc{33}{14.36} & \cc{89}{21.98} & \cc{90}{48.72} & \cc{30}{10.65} & \cc{67}{17.48} & \cc{88}{37.06} & \cc{13}{4.93} & \cc{50}{8.70} \\
Deepseek-V3.2 & Embedding & \cc{55}{32.08} & \cc{21}{3.62} & \cc{43}{6.50} & \cc{78}{40.48} & \cc{23}{9.99} & \cc{65}{16.02} & \cc{79}{42.74} & \cc{23}{8.40} & \cc{54}{14.04} & \cc{73}{30.83} & \cc{7}{2.81} & \cc{30}{5.15} \\
GLM-4.7 & Embedding & \cc{52}{30.19} & \cc{43}{7.51} & \cc{79}{12.03} & \cc{56}{29.06} & \cc{39}{17.04} & \cc{87}{21.48} & \cc{62}{33.33} & \cc{49}{17.73} & \cc{89}{23.15} & \cc{55}{23.16} & \cc{19}{7.03} & \cc{62}{10.78} \\
GPT-5.2 & Embedding & \cc{100}{58.49} & \cc{17}{2.93} & \cc{37}{5.58} & \cc{100}{51.62} & \cc{31}{13.43} & \cc{87}{21.31} & \cc{90}{48.72} & \cc{25}{8.85} & \cc{58}{14.98} & \cc{98}{41.05} & \cc{11}{4.18} & \cc{44}{7.59} \\
Qwen-480B-Coder & Embedding & \cc{39}{22.64} & \cc{59}{10.26} & \cc{92}{14.12} & \cc{51}{26.52} & \cc{41}{18.23} & \cc{88}{21.61} & \cc{51}{27.35} & \cc{47}{16.75} & \cc{80}{20.78} & \cc{51}{21.25} & \cc{16}{5.97} & \cc{54}{9.32} \\
\bottomrule
\end{tabular}%
}
\end{table}

Table~\ref{tab:model_performance_detailed} and Table~\ref{tab:model_comparison} shows additional statistics in experiment.

\section{Case Study} \label{app:case-study}

In this section, we present case studies illustrating correct and incorrect review comments generated by models under Agent-based and similarity-based retrieval settings. For each case, we provide the Repo, PR ID, File Path, along with the Diff Hunk in the file under review, the generated review comment, the retrieved code context (where applicable), and an analysis of the erroneous comments. We observe that current models continue to suffer from knowledge errors during code review. Furthermore, noisy data introduced by context retrieval is identified as a significant factor contributing to the generation of incorrect review comments.

\begin{figure}[hp]
    \centering
    \includegraphics[width=\linewidth]{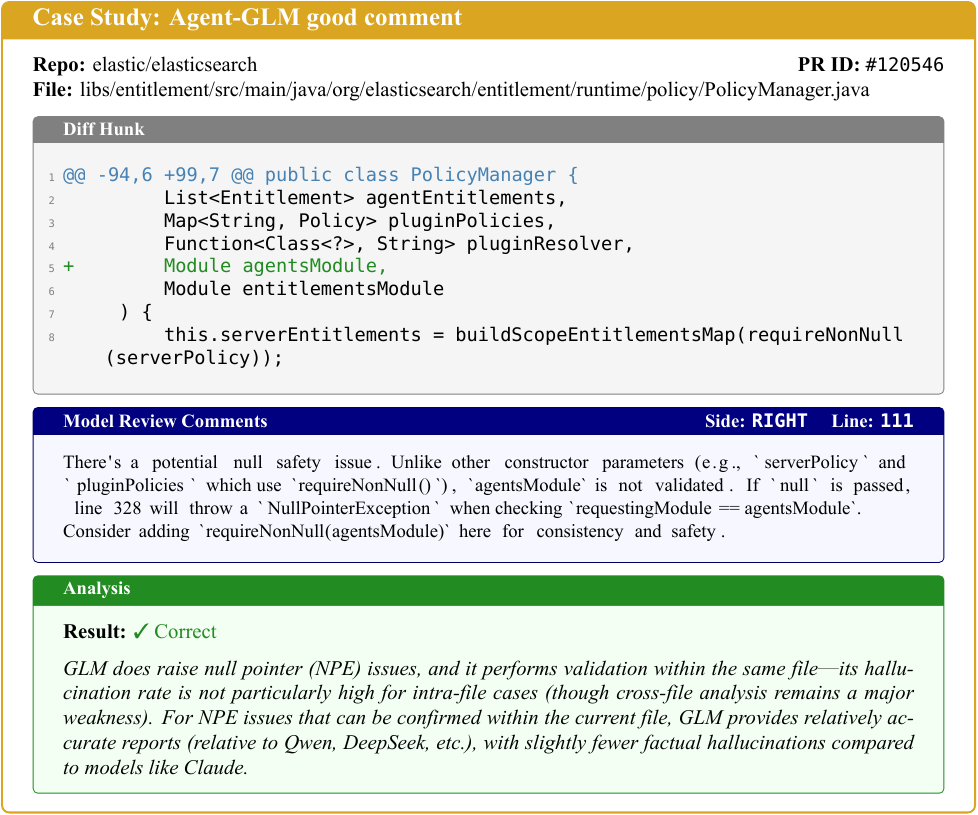}
    \caption{A Sample Correct Case Using Agent}
    \label{fig:case-1}
\end{figure}

\begin{figure}[hp]
    \centering
    \includegraphics[width=\linewidth]{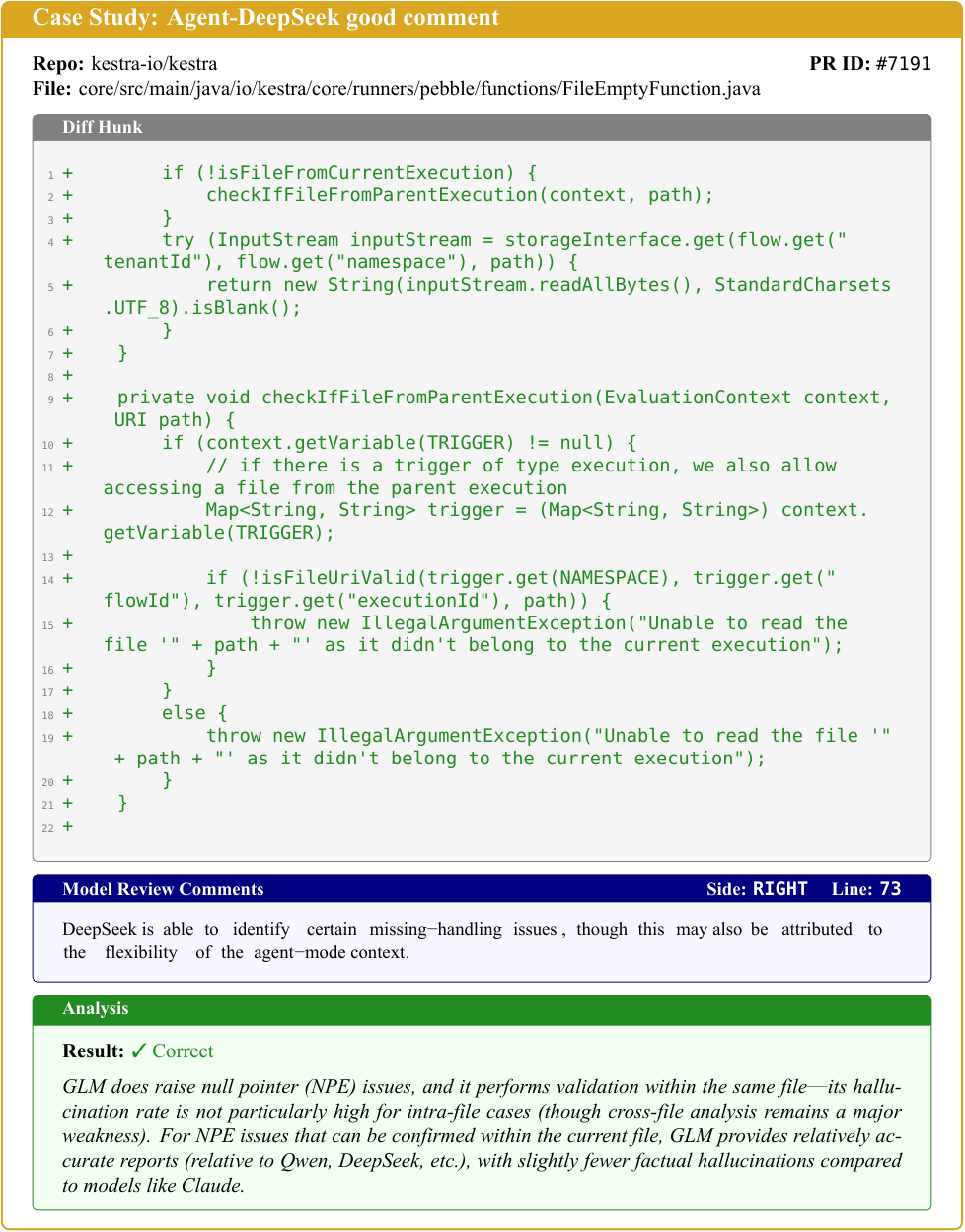}
    \caption{A Sample Correct Case Using Agent}
    \label{fig:case-2}
\end{figure}

\begin{figure}[hp]
    \centering
    \includegraphics[width=\linewidth]{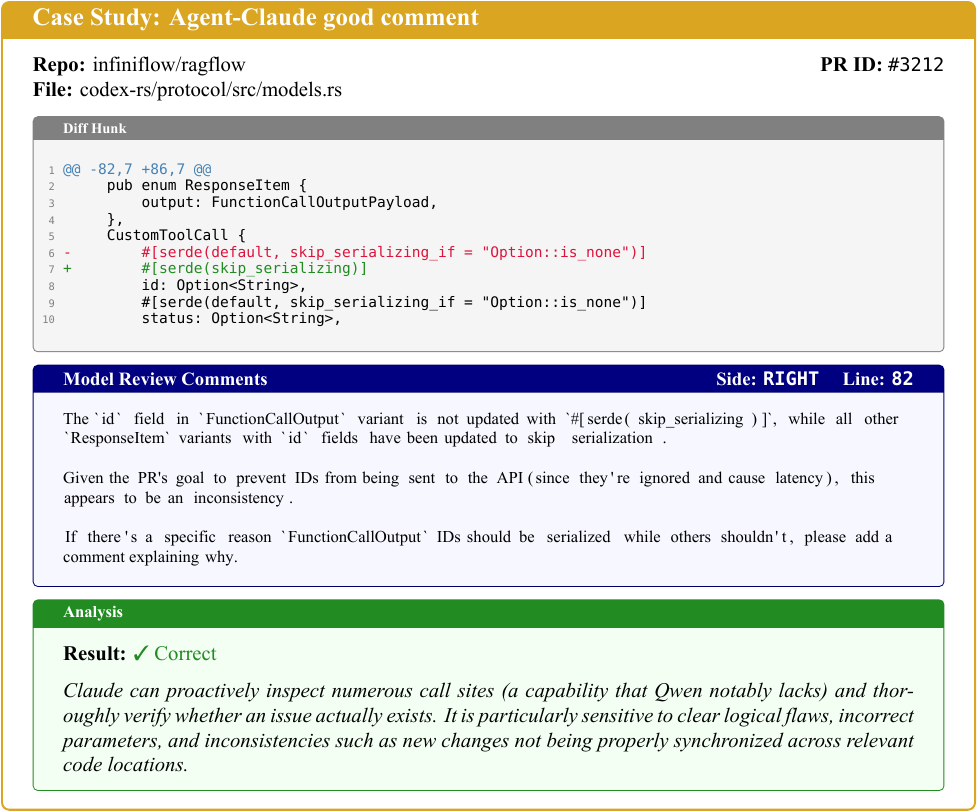}
    \caption{A Sample Correct Case Using Agent}
    \label{fig:case-3}
\end{figure}

\begin{figure}[hp]
    \centering
    \includegraphics[width=0.85\linewidth]{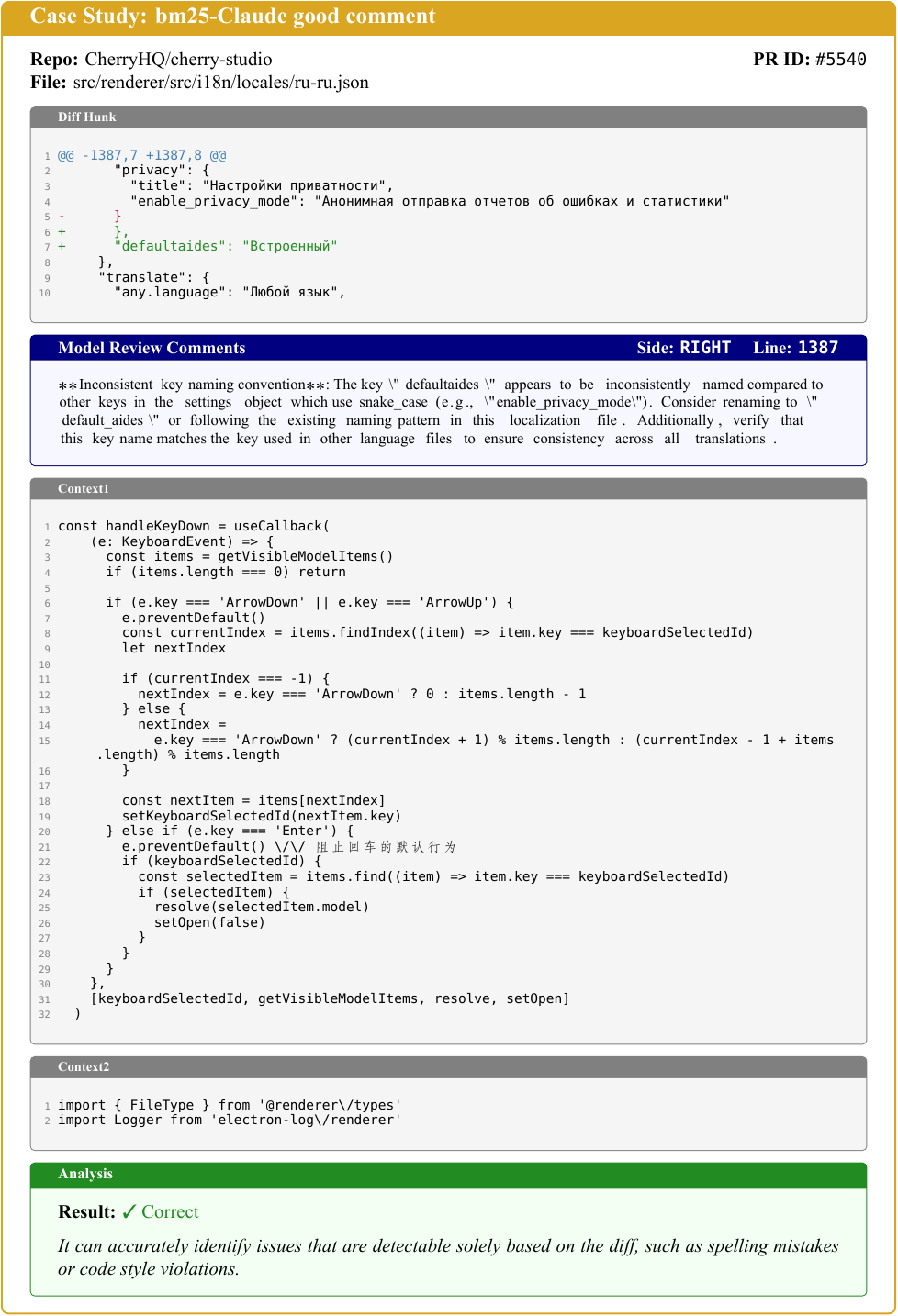}
    \caption{A Sample Correct Case Using BM25/Embedding Retrieval}
    \label{fig:case-4}
\end{figure}

\begin{figure}[hp]
    \centering
    \includegraphics[width=\linewidth]{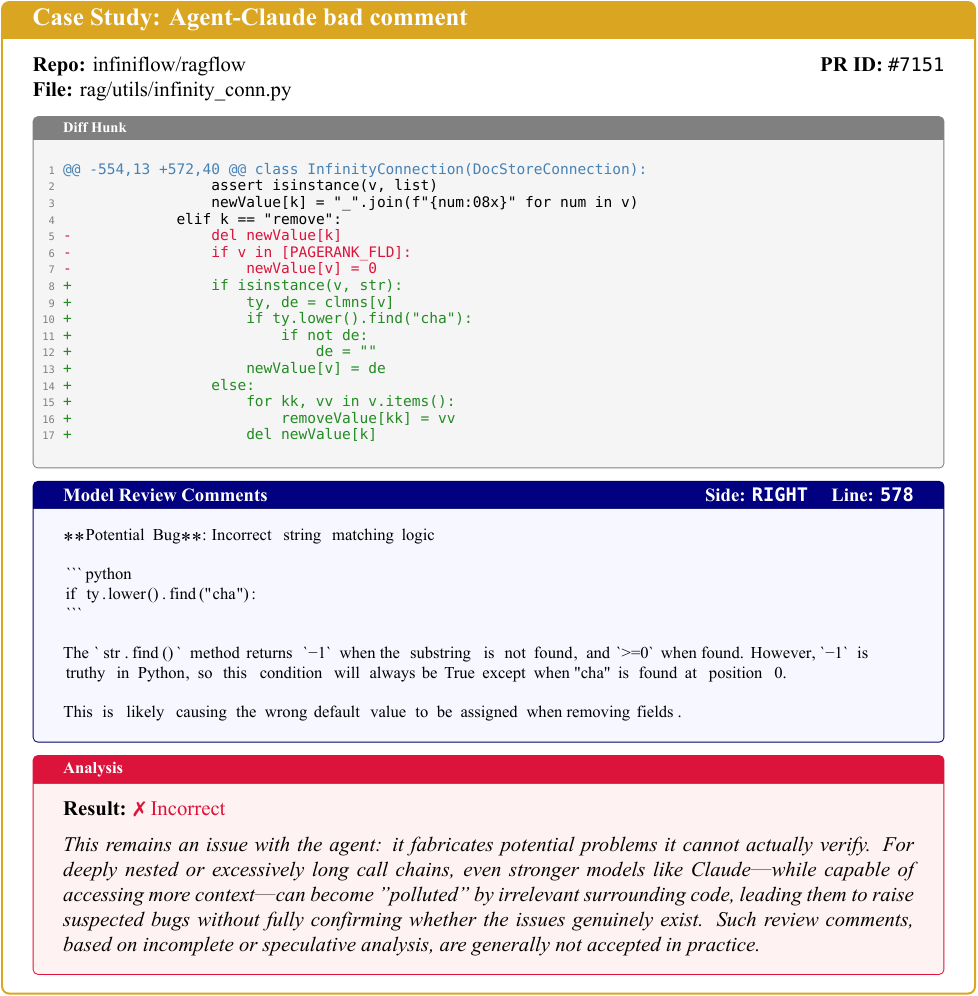}
    \caption{A Sample Error Case Using Agent }
    \label{fig:case-5}
\end{figure}

\begin{figure}[hp]
    \centering
    \includegraphics[width=\linewidth]{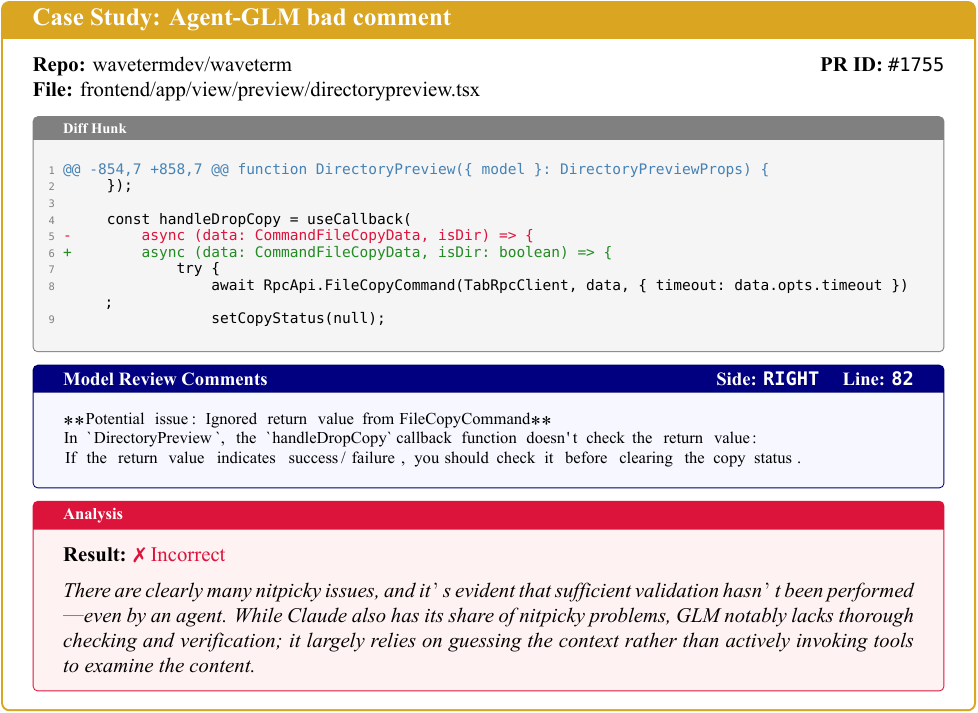}
    \caption{A Sample Error Case Using Agent}
    \label{fig:case-6}
\end{figure}

\begin{figure}[hp]
    \centering
    \includegraphics[width=0.95\linewidth]{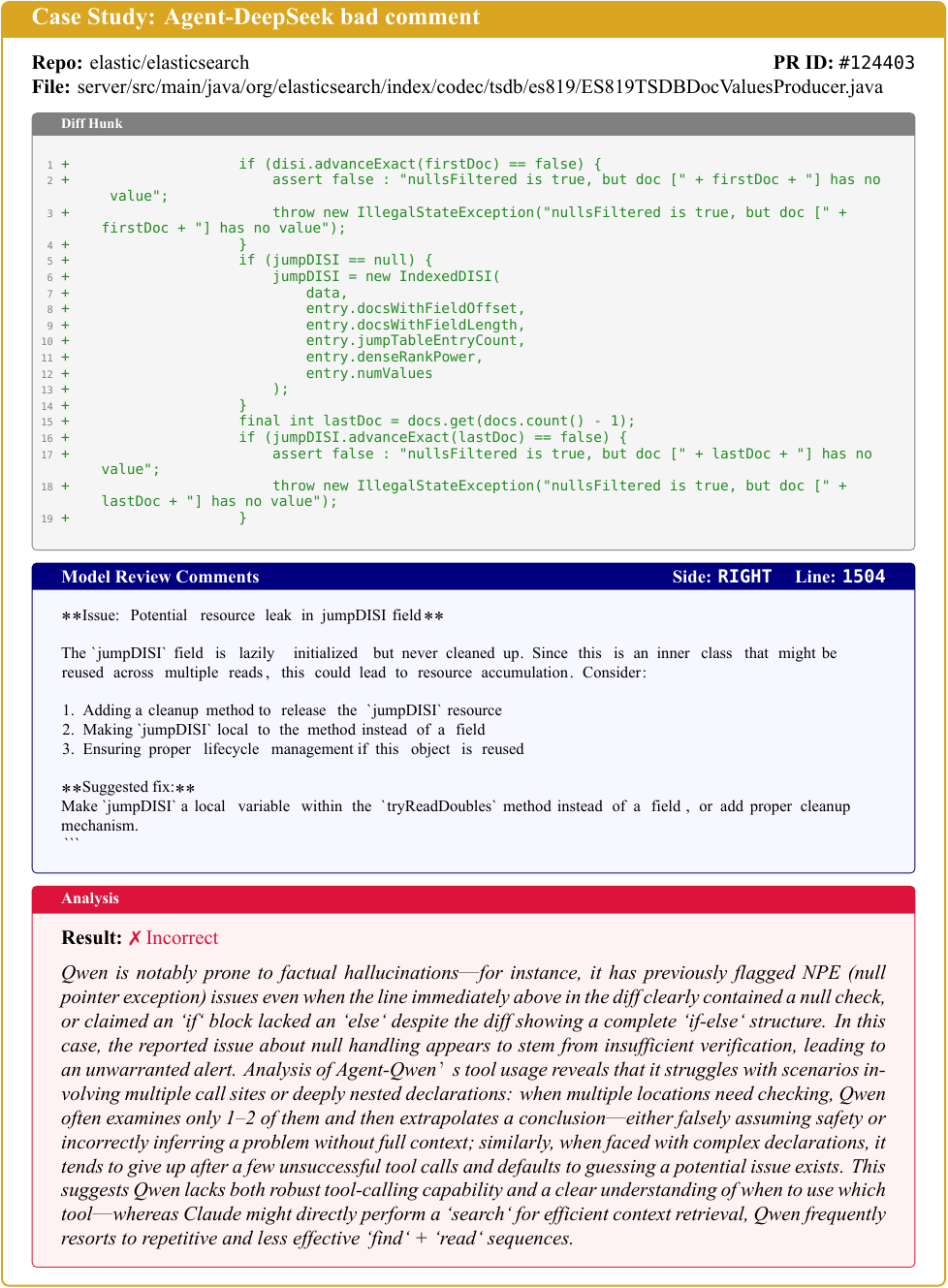}
    \caption{A Sample Error Case Using Agent}
    \label{fig:case-7}
\end{figure}

\begin{figure}[hp]
    \centering
    \includegraphics[width=0.95\linewidth]{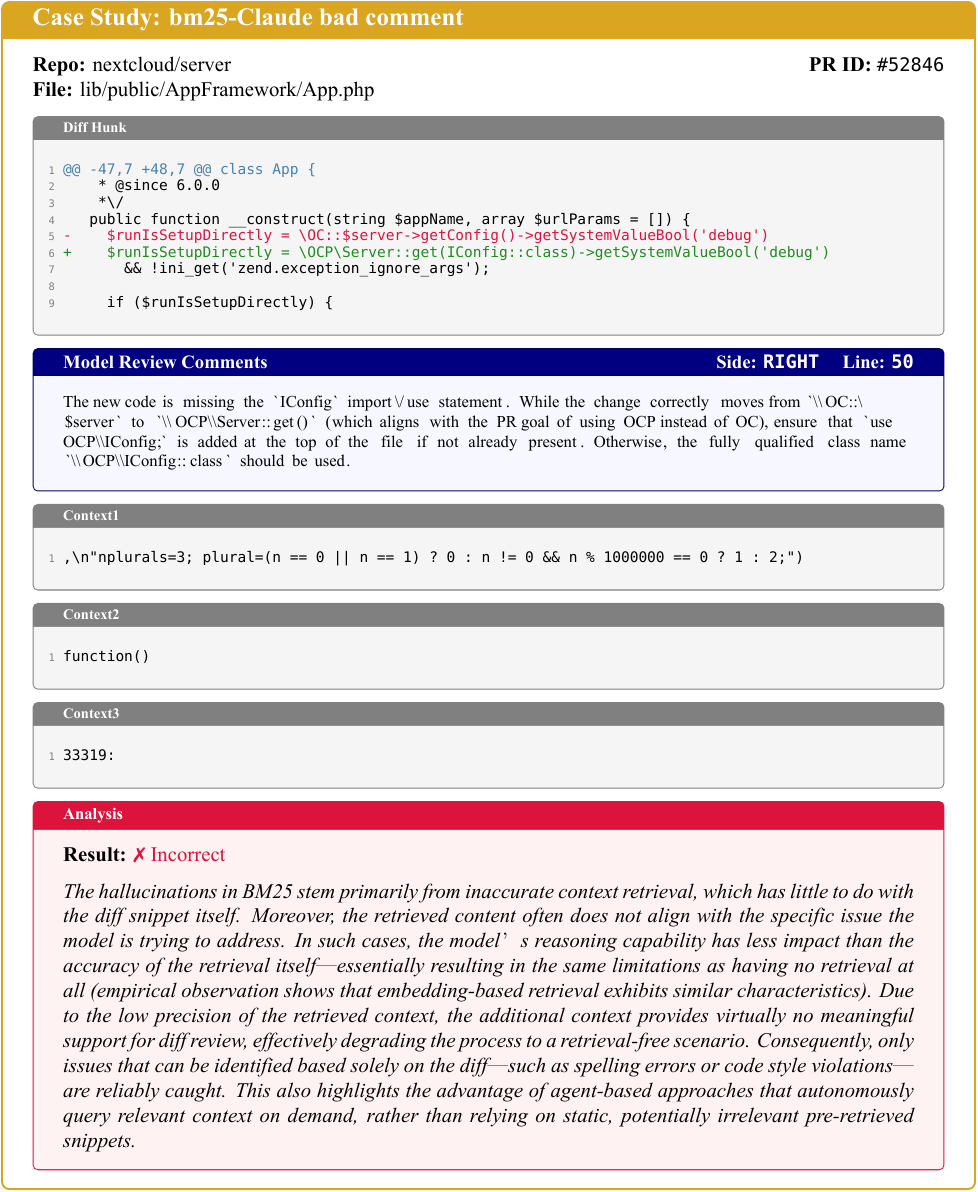}
    \caption{A Sample Error Case Using BM25/Embedding Retrieval}
    \label{fig:case-8}
\end{figure}

\begin{figure}[hp]
    \centering
    \includegraphics[width=0.7\linewidth]{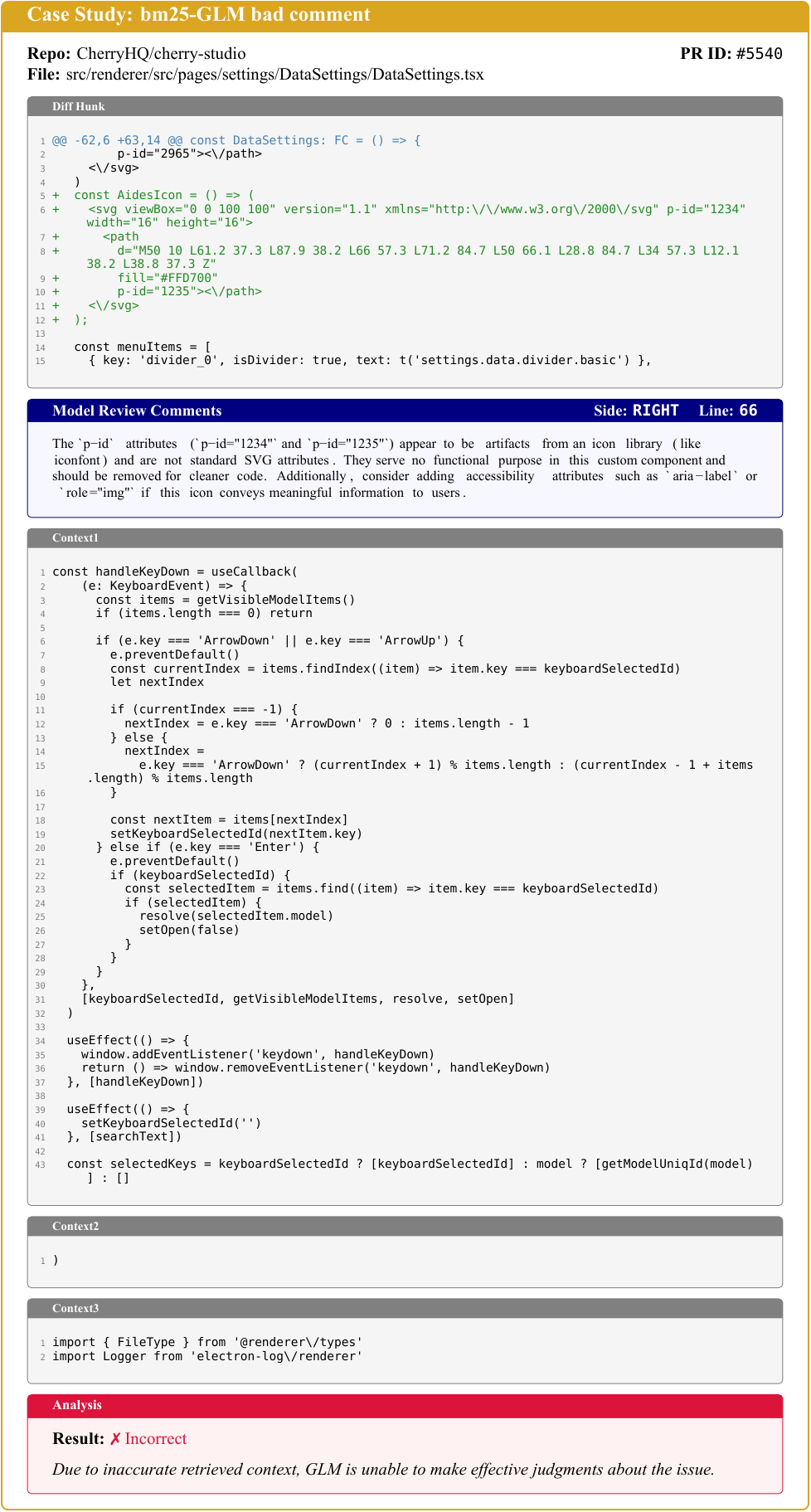}
    \caption{A Sample Error Case Using BM25/Embedding Retrieval}
    \label{fig:case-9}
\end{figure}

\begin{figure}[hp]
    \centering
    \includegraphics[width=0.8\linewidth]{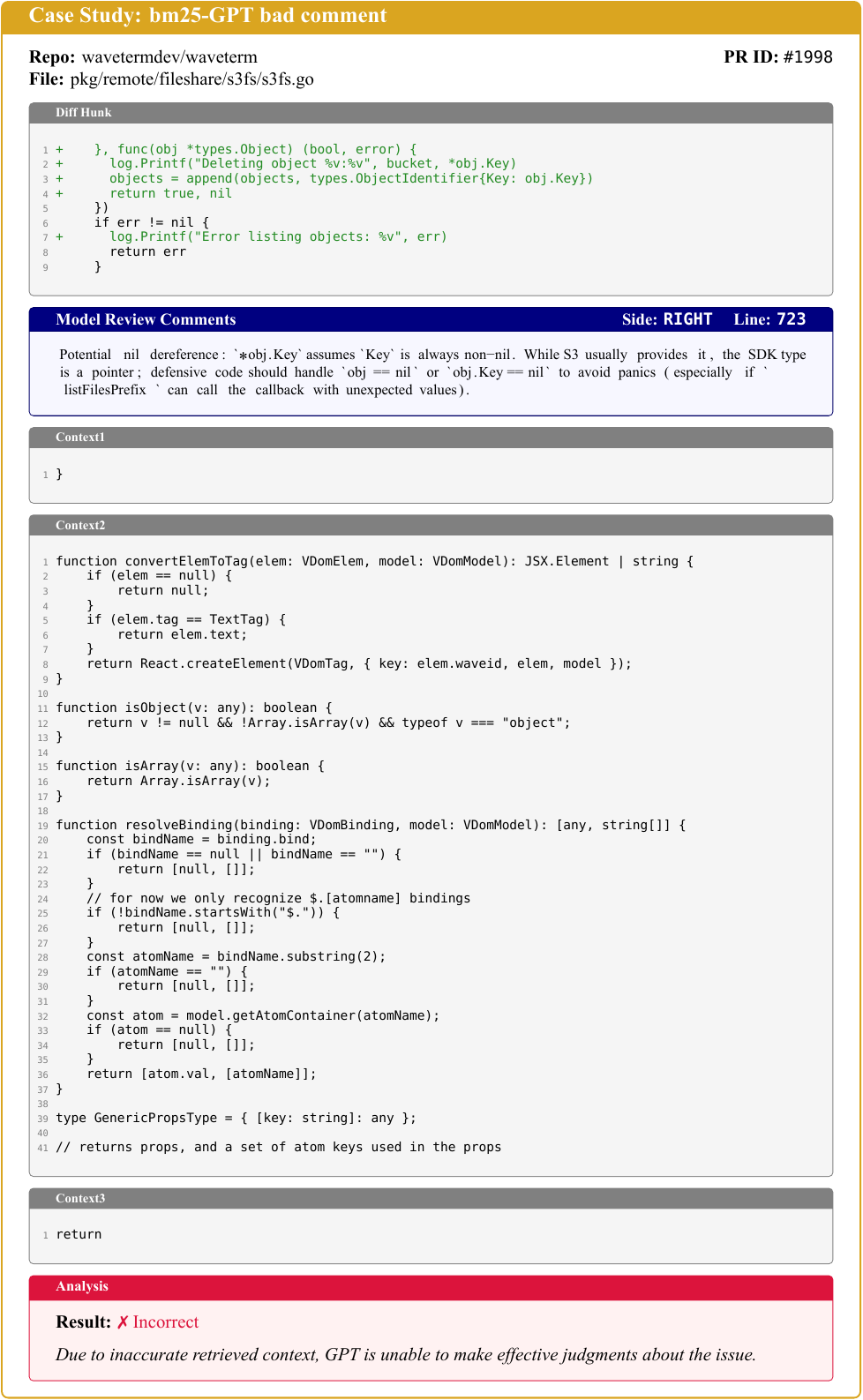}
    \caption{A Sample Error Case Using BM25/Embedding Retrieval}
    \label{fig:case-10}
\end{figure}

\begin{figure}[hp]
    \centering
    \includegraphics[width=0.75\linewidth]{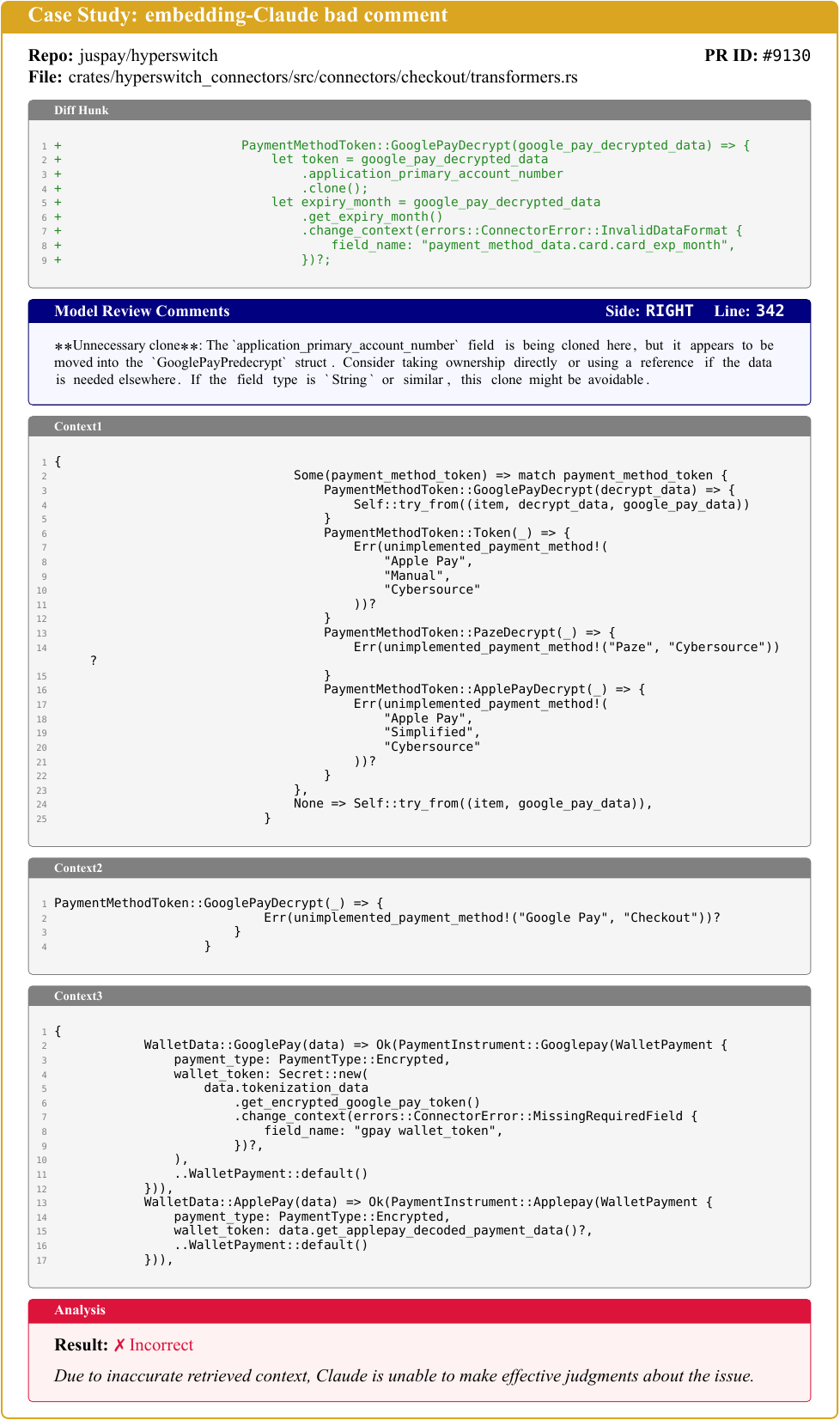}
    \caption{A Sample Error Case Using BM25/Embedding Retrieval}
    \label{fig:case-11}
\end{figure}

\end{document}